\documentclass[useAMS,usenatbib]{mn2e}

\def\msunyr{$\mathrm{M}_{\odot}\,\mathrm{yr}^{-1}$}
\def\kms{$\mathrm{km}\,\mathrm{s}^{-1}$}
\def \cm2{cm$^{-2}$}

\usepackage{subfigure}
\usepackage{graphicx}
\usepackage{wasysym}
\usepackage{stmaryrd}
\usepackage{txfonts}
\usepackage{natbib}
\usepackage{color}
\usepackage{colortbl}
\usepackage{multirow} 

\def   \aj {{\rm {AJ}}}
\def   \araa {{\rm {ARA\&A}}}
\def   \apj {{\rm {ApJ}}}

\def   \apjs {{\rm {ApJS}}}

\def   \aap {{\rm {A\&A}}}

\def   \aaps {{\rm {A\&AS}}}

\def   \mnras {{\rm {MNRAS}}}

\def   \jqsrt {{\rm {Journal of Quantitative Spectroscopy and Radiative Transfer}}}
\def   \apjl{\rm {ApJL}}
\def   \nat{\rm {Nat.}}

\title[CO bandhead emission of MYSOs]{CO bandhead emission of
massive young stellar objects: determining disc properties\thanks{Based on observations made with the ESO Very
Large Telescope at the Cerro Paranal Observatory under programme ID
079.C-0725}}

\author[J.~D.~Ilee et al.]
{\parbox{\textwidth}{J.~D.~Ilee$^{1}$\thanks{E-mail: \texttt{pyjdi@leeds.ac.uk}},
H.~E.~Wheelwright$^{2}$,
R.~D.~Oudmaijer$^{1}$,
W.~J.~de Wit$^{3}$, 
L.~T.~Maud$^{1}$,
M.~G.~Hoare$^{1}$,
S.~L.~Lumsden$^{1}$,
T.~J.~T.~Moore$^{4}$,
J.~S.~Urquhart$^{2}$ and
J.~C.~Mottram$^{5}$} \vspace{0.4cm} \\
\parbox{\textwidth}{
$^{1}$School of Physics and Astronomy, EC Stoner Building, University of Leeds, Leeds, LS2 9JT, UK\\
$^{2}$Max-Planck-Institut f\"{u}r Radioastronomie, Auf dem H\"{u}gel 69, 53121, Bonn, Germany\\
$^{3}$European Southern Observatory, Alonso de Cordova 3107, Vitacura, Santiago, Chile\\
$^{4}$Astrophysics Research Institute, Liverpool John Moores University, Twelve Quays House, Egerton Wharf, Birkenhead CH41 1LD\\
$^{5}$Leiden Observatory, Leiden University, PO Box 9513, 2300 RA Leiden, the Netherlands\\
}}

\begin{document}
\date{Accepted 2012 December 3.  Received 2012 November 21; in original form 2012 October 29}
\pagerange{\pageref{firstpage}--\pageref{lastpage}} \pubyear{2012}

\maketitle

\label{firstpage}

\begin{abstract}

Massive stars play an important role in many areas of astrophysics,
but numerous details regarding their formation remain unclear. In this
paper we present and analyse high resolution ($R \sim 30\,000$)
near--infrared $2.3\,\micron$ spectra of 20 massive young stellar
objects from the RMS database, in the largest such study of CO first
overtone bandhead emission to date. We fit the emission under the
assumption it originates from a circumstellar disc in Keplerian
rotation.  We explore three approaches to modelling the physical
conditions within the disc - a disc heated mainly via irradiation from
the central star, a disc heated mainly via viscosity, and a disc in
which the temperature and density are described analytically.  We find
that the models described by heating mechanisms are inappropriate
because they do not provide good fits to the CO emission spectra.  We
therefore restrict our analysis to the analytic model, and obtain good
fits to all objects that possess sufficiently strong CO emission,
suggesting circumstellar discs are the source of this emission.  On
average, the temperature and density structure of the discs correspond
to geometrically thin discs, spread across a wide range of
inclinations.  Essentially all the discs are located within the dust
sublimation radius, providing strong evidence that the CO emission
originates close to the central protostar, on astronomical unit
scales.  In addition, we show that the objects in our sample appear no
different to the general population of MYSOs in the RMS database,
based on their near-- and mid--infrared colours. The combination of
observations of a large sample of MYSOs with CO bandhead emission and
our detailed modelling provide compelling evidence of the presence of
small scale gaseous discs around such objects, supporting the scenario
in which massive stars form via disc accretion.

\end{abstract}


\section{Introduction}

Massive stars are important from stellar to galactic scales. Their
high temperature and luminosity results in the injection of large
amounts of ionizing radiation and kinetic energy into their
surroundings, which shapes the local interstellar medium (ISM) and may
trigger nearby star formation.  They also deposit chemically enriched
material into the ISM via continuous mass loss during their lifetime
and in supernova explosions.  However, the formation mechanisms of
massive stars (M $\gtrsim$ 8\,M$_{\odot}$) are poorly understood
\citep[see the review of][]{zinnecker_2007}.  Given their importance
in stellar and galactic evolution, it is crucial to understand how
they form.

\smallskip

The precursors of massive stars, massive young stellar objects
(MYSOs), possess a short Kelvin--Helmholtz contraction timescale
($10^{4-5}$\,yr, \citealt{mottram_2011b}) and thus reach the main
sequence and obtain a high luminosity while still enshrouded in their
natal cloud material.  This high luminosity presents a challenge to
theories of massive star formation, particularly when considering a
scaled up version of low mass star formation (as in
\citealt{shu_1987}, see \citealt{norberg_2000}). There is only a short
time for the star to accumulate sufficient mass before the protostar
ionises the surrounding material, which, along with significant
radiation pressure from the protostar, may halt further accretion
\citep{larson_1971,kahn_1974,wolfire_1987}. However, recent 3D
hydrodynamic models indicate that discs circumvent the proposed
barrier to massive star formation by facilitating the accretion of
matter on to the central object
\citep{krumholz_2009,kuiper_2010,kuiper_2011}.  In addition,
observations have shown that some ultra-compact H\,{\sc ii} regions
have outflows, usually associated with ongoing accretion
\citep[e.g.][]{klaassen_2011}. Thus, continued accretion must be
possible after the star reaches the main-sequence and begins ionizing
its surroundings.

\smallskip 

Confirming the presence of discs around MYSOs presents a considerable
observational challenge. Such objects are relatively rare, and still
embedded in molecular cloud material, making them optically
invisible. There have been a handful of detections of discs around
MYSOs \citep[see][]{patel_2005, jiminez_2007, kraus_2010,
carrasco_2012} but the disc properties are difficult to determine.
Observations at longer wavelengths (such as the far infrared and
submillimetre) only probe disc properties at large distances from the
central protostar. Furthermore, very few studies can be conducted with
sufficient angular resolution to probe astronomical unit sized scales,
which is necessary to study the inner regions of circumstellar
discs. The exception is the observation of the MYSO
G310.0135$+$00.3892 with the Very Large Telescope Interferometer
(VLTI) and AMBER \citep{petrov_2007} reported by \citet{kraus_2010},
which achieved a maximum resolution of approximately 10 au. This
provided unique information on the geometry of the $K-$band continuum
emitting material. However, this study involved only a single
object. Therefore, observations of a large sample of MYSOs using an
alternative method that can probe close to the central protostar are
required.

\smallskip

The inner regions of gaseous discs are difficult to access
observationally, especially as near-infrared (NIR) interferometric
studies are limited to isolated objects which are bright in the
near-infrared \citep{tatulli_2008,wheelwright_2012_amber}.  Therefore,
to study the inner discs of MYSOs, we must employ indirect methods.
The CO molecule is an ideal tracer of these regions because the
coupled rotational and vibrational excitation causes a distinctive
emission feature, the CO bandhead, so called because they appear in
bands in low-resolution spectra.  The first overtone $v=$ 2--0
bandhead emission at $2.3\,\micron$ occurs in warm ($T =$
2500--5000\,K) and dense (n $>$ $10^{11}$\,cm$^{-3}$) gas.  These are
the conditions expected in the inner regions of accretion discs. This
makes CO bandhead emission a valuable tool that allows us to trace
these regions. In addition, because this feature is the result of
transitions across a range of energy levels (and therefore,
temperatures), it also allows us to probe the physical properties
throughout the disc.  Previous studies of CO bandhead emission have
been successful in fitting spectra of young stars with a range of
masses under the assumption that the emission originates from a
circumstellar disc \citep{carr_1989, chandler_1995, bik_2004,
blum_2004, davies_2010, wheelwright_2010}, but a study involving a
significant number of MYSOs has yet to be performed.

\smallskip

This has been partly due to a lack of a representative sample of
MYSOs. Early searches for MYSOs were conducted using the \textit{IRAS}
point source catalogue \citep{molinari_1996,sridharan_2002}. This
suffered from source confusion due to the large beam size
(2--5$\,$arcmin at $100\,\micron$) and were biased to isolated objects
away from the Galactic plane.  This issue has been addressed by the
Red \textit{MSX} Source (RMS) survey \citep{lumsden_2002}, which is an
unbiased survey of MYSOs throughout the Galaxy. It is drawn from the
\textit{MSX} mid--infrared survey of the Galactic plane
\citep{egan_2003}, which has a resolution of 18 arcsec, allowing
detection of sources in previously unresolved regions.  An extensive
multi-wavelength campaign has been conducted to identify contaminant
objects such as ultra-compact H\,{\sc ii} regions and planetary nebula
\citep{urquhart_2007a, urquhart_2009a, mottram_2007, mottram_2010},
finally yielding approximately 500 candidate MYSOs in the database.
Kinematic distance estimates to the MYSOs were obtained from molecular
line observations \citep{urquhart_2007b, urquhart_2008,
urquhart_2009b,urquhart_2011,urquhart_2012}, while bolometric
luminosities have been determined from fits to the MYSOs' spectral
energy distributions (SEDs; \citealt{mottram_2011a}).

\smallskip

In this paper we study a subset of the RMS
database\footnote{http://www.ast.leeds.ac.uk/RMS/}. We utilise our
extensive low resolution spectroscopic survey of RMS sources
\citep[see, e.g.\ ][]{clarke_phd_2007, cooper_prep} to select objects
for a high resolution spectroscopic study of CO bandhead emission in
massive young stellar objects.  We detect CO emission in 20 MYSOs (and
two non-MYSOs), which is compared to kinematic models to assess
whether the emission originates in circumstellar discs. Finally, we
attempt to determine the properties of the CO emitting gas and
constrain the accretion rates of these objects.  Section \ref{sec:obs}
outlines the observations we have performed while Section
\ref{sec:discs} details our modelling.  Section \ref{sec:results}
presents our observations and model fits to the spectra, along with an
analysis of the best fitting parameters.  Section \ref{sec:discussion}
discusses our findings and Section \ref{sec:conclusions} presents our
conclusions from this work.

\section{Observations and sample selection}
\label{sec:obs}

Table \ref{tab:obs} presents the observational parameters of the 20
MYSO, and two non-MYSO targets in our study that possessed CO
emission.  The data were taken using the CRIRES near--infrared
cryogenic spectrograph \citep{kaeuful_2004} on the Very Large
Telescope (VLT) over three nights in June 2007.  A spectral resolution
of $R \sim 30\,000$ was achieved ($\Delta \lambda=0.08\,\mathrm{nm}$
at $\lambda = 2.3\,\micron$) using a slit width of 0.6 arcsec.
Standard ABBA nodding along the slit was used.  The seeing conditions
varied from 0.8 to 1.2 arcsec.  A single pixel element represents
0.011$\,\mathrm{nm}$, while a resolution element covers approximately
seven pixels.  Using a central wavelength of $\lambda_{\mathrm{c}} =
2.286\,\micron$, the CO emission spans chips three and four of the
detector.  The CO bandhead peak is located on the third detector chip.
Telluric standard stars of spectral type A, featureless in the
wavelength range of interest, were observed between science targets
using an identical instrumental set-up and at similar airmasses to
ensure similar sky conditions.

\smallskip

The data were reduced using the ESO provided CRIRES pipeline, via the
\textsc{gasgano} data organiser (version 2.2.7).  Dark current was
removed and detector linearity corrections applied, then master flat
frames and bad pixel maps were used to correct the spectra.  The final
spectra were extracted using the optimal extraction algorithm.
Wavelength calibration was performed using a cross correlation with
the \textsc{hitran} model catalogue \citep{rothman_1998} and OH lines.
The standard stars were reduced and extracted in the same manner.  The
final spectra were obtained by division of the object spectra by their
corresponding standard star to remove telluric spectral features.  To
further remove the effect of bad pixels, the spectra were cleaned
using a sigma-clipping algorithm that removed any pixels with a value
above three times the standard deviation of the pre-bandhead portion
of the spectrum.  This was determined to be the maximum amount of
cleaning that could be performed without affecting the appearance of
real spectral features.

\smallskip

\begin{center}
  \begin{table*}
    \begin{center}
      \begin{minipage}{\textwidth}
	\begin{center}
	  \renewcommand{\footnoterule}

	  \caption{Observed quantities for each object.  The
	  bolometric luminosity $L_{\mathrm{Bol}}$, 2MASS $K-$band
	  magnitude and distance $d_{\mathrm{kin}}$ are all taken from
	  the RMS database unless otherwise stated.}
          \label{tab:obs}
           \begin{tabular}{llllllllllll}

\hline
Object                          	&	RA	        &   	Dec.         &    $L_{\mathrm{Bol}}$	&  $K$       &  $d_{\mathrm{kin}}$  & $t_{\mathrm{int}}$  &   Blue           & Chip  & S/N & CO                        \\
                                        &       (J2000)         &        (J2000)         &      ($\mathrm{L}_{\odot}$)  &(mags)    &    (kpc)             &            (h)      &  Shoulder             & Four  &    & $\sigma$                        \\
\hline
\multicolumn{3}{l}{\textbf{MYSOs}}\\
G012.9090$-$00.2607     &	18:14:39.56 	&       $-$17:52:02.3	&	$5.4\times10^{4}$	&	9.2	&	3.8	&	5.0	  &Y	&	Y	&63	&12 		\\
G033.3891$+$00.1989	&	18:51:33.82 	&       $+$00:29:51.0	&	$1.0\times10^{4}$	&	7.2	&	5.1	&	0.3	  &N	&	Y	&81	&10 		\\
G035.1979$-$00.7427	&	18:58:12.99 	&       $+$01:40:31.2   &	$3.1\times10^{4}$	&	12.6	&	2.2	&	1.6	  &N	&	N	&18	&7  		\\
G270.8247$-$01.1112	&       09:10:30.89 	&       $-$49:41:29.8 	&	$9.9\times10^{3}$	&	10.1	&	7.7	&	8.3	  &Y	&	Y	&61	&11		\\
G282.2988$-$00.7769	&       10:10:00.32     &       $-$57:02:07.3   &	$8.9\times10^{3}$	&	7.0	&	5.5	&	0.1	  &Y	&	Y	&114	&11	        \\
G287.3716$+$00.6444	&	10:48:04.55 	&       $-$58:27:01.5  	&	$2.8\times10^{4}$	&	7.5	&	5.6	&	0.3	  &N	&	N	&128	&12		\\
G293.8947$-$00.7825	&	11:32:32.82     &       $-$62:15:43.1   &	$1.2\times10^{4}$	&	8.8	&	10	&	2.1	  &N	&	Y	&102	&6		\\
G296.2654$-$00.3901	&	11:53:10.93 	&       $-$62:30:20.0	&	$4.5\times10^{3}$	&	8.9	&	8.9	&	2.1	  &N	&	Y	&58	&8		\\
G305.2017$+$00.2072	&	13:11:10.45 	&       $-$62:34:38.6	&	$4.9\times10^{4}$	&	9.4	&	3.6	&	2.1	  &N	&	Y	&63	&18		\\
G308.9176$+$00.1231	&	13:43:01.70	&	$-$62:08:51.2	&	$1.4\times10^{5}$	&	6.4	&	4.6	&	0.1	&Y	&	N	&64	&7		\\
G310.0135$+$00.3892 	&	13:51:37.85	&	$-$61:39:07.5	&	$5.7\times10^{4}$	&	4.9	&	3.3	&	0.1	&Y	&	N	&120	&8		\\
G332.0939$-$00.4206	&	16:16:16.46	&	$-$51:18:25.2	&	$1.8\times10^{5}$	&	5.9	&	3.5	&	0.1	&-	&	N	&58	&5		\\
G332.9868$-$00.4871	&	16:20:37.81	&	$-$50:43:49.6	&	$2.6\times10^{4}$	&	9.3	&	3.5	&	3.3	&Y	&	N	&47	&10		\\
G338.9377$-$00.4890	&	16:45:08.23	&	$-$46:22:18.5	&	$2.1\times10^{3}$	&	9.1	&	3.2	&	2.1	&Y	&	N	&50	&7		\\
G339.6816$-$01.2058	&	16:51:05.95	&	$-$46:15:52.4	&	$1.6\times10^{4}$	&	8.5	&	2.6	&	0.3	&-	&	N	&74	&4		\\
G347.0775$-$00.3927	&	17:12:25.81	&	$-$39:55:19.9	&	$1.8\times10^{3}$	&	8.5	&	1.6	&	2.1	&Y	&	Y	&62	&7		\\
IRAS08576$-$4334$^{\mathrm{\S}}$	&	08:59:27.40	&	$-$43:45:03.7	&	-	&	9.4	&	2.2$^{\mathrm{\|}}$	&	5.0	&N	&	Y	&41	&40		\\
IRAS16164$-$5046$^{\mathrm{\S}}$	&	16:20:11.06	&	$-$50:53:16.2	&	$1.0\times10^{5}$$^{\mathrm{\dag}}$	&	9.5$^{\mathrm{\dag}}$	&	3.6$^{\mathrm{\dag}}$	&5.0		&Y	&	Y	&61	&14	\\
IRAS17441$-$2910$^{\mathrm{\S}}$	&	17:44:09.60	&	$-$29:10:58.0	&	$5.2\times10^{5}$$^{\mathrm{\ast}}$	&	5.3	&	9.8$^{\mathrm{\ast}}$	&	0.4		&Y	&	Y	&	33&9	\\
M8E-IR$^{\mathrm{\S}}$           	&	18:04:53.26	&	$-$24:26:42.3	&	-	               &	4.4	&	1.9$^{\mathrm{\ddag}}$	&	0.2			&    N	& Y	&125	&12		\\
\hline
\multicolumn{3}{l}{\textbf{Non-MYSOs} (see Appendix \ref{sec:others})}\\
G332.9457$+$02.3855	&	16:08:12.08	&	$-$48:40:38.5	&	860	&	9.7	&	1.9	&	0.04	&N	&	Y	&36	&8		\\
G338.5459$+$02.1175	&	16:32:32.19	&	$-$44:55:30.6	&	16	&	7.2	&	0.3	&	5.0	&N	&	Y	&60	&17		\\
\hline

  \end{tabular} 
	\end{center}
        \small{$\mathrm{\S}$: Object is not a member of the RMS database, but was selected based on previously observed CO emission.} \\
	\small{$\mathrm{\dag}$: From \citet{bik_2006}.}\\
        \small{$\mathrm{\ddag}$: From \citet{chini_1981}.}\\
	\small{$\mathrm{\ast}$: From \citet{walsh_1997}.}\\
        \small{$\mathrm{\|}$: Kinematic distance from the rotation curve of \citet{brand_1993} and $v_{\mathrm{LSR}}$ = 7.5\,\kms \citep{bronfman_1996}.}\\
 \end{minipage}
    \end{center}
  \end{table*}
\end{center}

\subsection{Observational Results}

The spectra, presented in Section \ref{sec:results}, exhibit a range
of bandhead shapes and strengths.  Ten objects possess a so-called
`blue shoulder', in which there is prominent emission on the shorter
wavelength side of the bandhead peak.  This is caused by
doppler-shifted rotational lines, and may be indicative of rotational
motion in a circumstellar disc, or an outflowing wind
\citep{kraus_2000}.  The other ten objects show steep rises in the
bandhead slope.  Several objects (for instance, G270.8247$-$01.1112)
possess a residual telluric feature in the pre-bandhead portion of the
spectrum, but this did not affect our analysis.

\smallskip

The signal to noise ratio of the spectra ranges from approximately
20--150. The average bolometric luminosity of the objects is $5\times
10^{4}\,\mathrm{L}_{\odot}$, typical of other objects within the RMS
Survey \citep{mottram_2011b}.  The objects are generally bright in the
$K-$band, but this is a selection effect based on observational
requirements.

\smallskip

The objects G332.9457$+$02.3855 and G338.5459$+$02.1175 were
originally classified as MYSOs, but since the observation date have
been found to have too low a luminosity for this to be the case, and
are likely lower mass young stellar objects.  However, since they both
possess strong CO emission we have included them in the final sample
and discuss them in Appendix \ref{sec:others}.

\smallskip

\begin{figure}
\centering

\includegraphics[width=1.0\columnwidth]{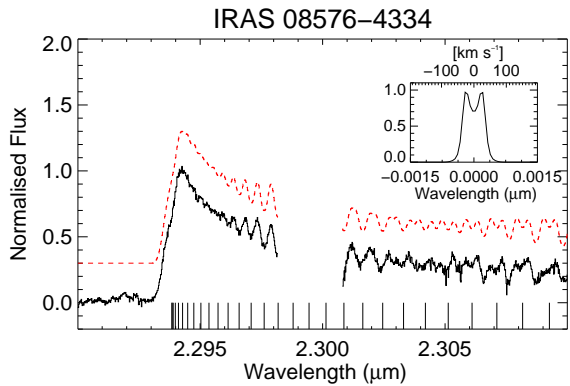}

\caption{Spectrum of the first overtone bandhead emission of IRAS
08576$-$4334 (solid black).  The gap in the data is due to the spacing
between chips three (left) and four (right) of the detector.  The best
fitting model is shown above (dashed red) and has been shifted upwards
for clarity.  The theoretical wavelengths of the transitions that make
up the bandhead are marked on the abscissa, and have been shifted to
account for relative motion.  The panel shows the line shape of the
$J=51$-50 transition that was adopted for the best fitting model.  }
\label{fig:iras08}
\end{figure}

Figure \ref{fig:iras08} shows the bandhead of IRAS 08576$-$4334, which
is a good example of the CO bandhead feature due to the prominent
emission and the high signal-to-noise ratio.  Plotted above the data
is our final best fitting model (discussed in Section
\ref{sec:discs}).  The inset shows the line shape of the $J=51$--50
transition that is the result of the best fitting model, displaying a
double peaked profile due to the motion in the disc.

\section{Modelling the emission}
\label{sec:discs}

To determine whether the CO emission originates from circumstellar
discs and to constrain the physical properties of the emitting region,
we compare the observations to a model of emission from a disc.  There
are several ways in which the physical parameters within the disc can
be modelled.  For example, \citet{bik_2004} assume a simple isothermal
disc, \citet{berthoud_2007} assume the surface density and temperature
of the disc decrease as power laws.  \citet{carr_1989} and
\citet{chandler_1995} determine the temperature and density structure
by balancing the heating and cooling mechanisms at work on the surface
of a thin alpha disc \citep[see][]{shakura_1973} which is isothermal
in the vertical direction.

\smallskip

To fit the emission using a disc model, we first divide the disc into
75 radial rings each with 75 azimuthal cells. Each cell is assigned a
temperature and surface density calculated from the relevant disc
description (see Section \ref{sec:discmodels}).  We assume Keplerian
rotation for the disc, and the velocity of each cell to the
line-of-sight is determined assuming the disc is at an inclination $i$
to the vertical.

\smallskip

Our model of the CO emission is based on \citet{wheelwright_2010},
which was in turn based on that of \citet{kraus_2000}, and briefly
described here.  The population of the CO rotational levels, to a
maximum of $J=100$, for the $v=$ 2--0 vibrational transition is
determined in each cell according to the Boltzmann distribution, which
assumes local thermodynamic equilibrium, and a CO/H$_{2}$ ratio of
10$^{-4}$.  The intrinsic line shape of each transition is assumed to
be Gaussian, with a line width of $\Delta \nu$.  The intensity of
emission is then calculated from the equation
\begin{equation}
I_{\nu} = B_{\nu}(T)\left(1-e^{-\tau_{\nu}}\right).
\end{equation}

\smallskip

The emission from each cell is weighted by its solid angle subtended
by the cell on the sky, and wavelength shifted with respect to the
line-of-sight velocity due to the rotational velocity of the disc.
The emission from all cells is then summed together, smoothed to the
instrumental resolution, and shifted in wavelength to account for the
radial velocity of the object to produce the entire CO bandhead
profile for the disc.

\subsection{Disc models}
\label{sec:discmodels}

We have investigated the use of three different approaches to
model the disc. Here, we discuss their differences.

\subsubsection{Model A}

Our first model, model A, is purely analytic in nature
\citep[as in][]{berthoud_2007} and describes the excitation temperature and
surface number density as decreasing power laws,
\begin{eqnarray}
T(r) = T_{\mathrm{i}} \left( \frac{r}{R_{\mathrm{i}}} \right)^{p} \\
N(r) = N_{\mathrm{i}} \left( \frac{r}{R_{\mathrm{i}}} \right)^{q},
\end{eqnarray}
where $T_{\mathrm{i}}$ and $N_{\mathrm{i}}$ are the excitation
temperature and surface density at the inner edge of the disc
$R_{\mathrm{i}}$, and $p$ and $q$ are the exponents describing the
temperature and surface density gradient, respectively.  The optical
depth, $\tau$, is taken to be the product of the absorption
coefficient per CO molecule, and the CO column density.  Since we are
considering a geometrically thin disc, the column density is given by
the surface number density $N$.

\subsubsection{Model B}

Our second model, model B, is based on balancing the heating and
cooling descriptions of a disc as in \citet{carr_1989}, later updated
by \citet{chandler_1995}.  The disc is assumed to be dominated by
stellar heating, with a contribution from heating via viscosity, and
thus the temperature considered is the surface temperature of the
disc.  The disc is assumed to be steady state, and thus has a constant
accretion rate throughout.  The disc is under local thermodynamic
equilibrium.  Finally, the disc is isothermal in the vertical
direction, and is heated via absorption of the stellar radiation field
and the conversion of gravitational potential energy into thermal
energy via viscosity. The viscosity within the disc is described
by the alpha prescription \citep{shakura_1973}.  The mass accretion
rate of MYSOs is expected to be high (up to $10^{-3}$\,\msunyr) and
the discs are likely turbulent, therefore we adopt a value of
$\alpha=1.0$, corresponding to a highly viscous disc environment.

\smallskip

To obtain the excitation temperature and surface number density at a
certain radius within the disc, we balance the heating and cooling
rates at the disc surface. This condition is then iterated through
with increasing temperatures until fulfilled, and the final physical
properties are adopted for the particular radius. This is repeated
throughout the disc.

\subsubsection{Model C}

Our third model, model C, is again based on balancing the heating and
cooling mechanisms of a disc, but as described in \citet{vaidya_2009}.
In this disc, the heating via viscosity is assumed to dominate the
temperature structure, and as such the temperature here can be
considered the midplane temperature of the disc.  The disc is steady
state and we again adopt $\alpha=1.0$.

\smallskip

We again balance the heating and cooling rates, which provides the
temperature and number density at a specific radius within the disc
via iteration.  The heating due to irradiation from the star is not
included in this iteration because we are considering the midplane
temperature of the disc, but is instead added to the temperature
reported from this convergence, yielding the final excitation
temperature.

\smallskip

In both model B and C, the mass accretion rate directly sets the
temperature structure throughout the disc.  The inner edge of the
emission region is set to the radius at which the temperature reaches
$5000\,$K (where we assume CO is destroyed), and the outer edge of the
disc is set to where the temperature drops below $1000\,$K (where we
assume CO is no longer ro-vibrationally excited).  The optical depth
is taken to be $\tau = \kappa \rho H$ where $H$ is the scale height of
the disc, and $\kappa$ is the opacity taken from
\citet{ferguson_2005}.

\begin{center}
\begin{table}
\begin{center}
\caption{Allowed ranges of parameters for the model fitting procedure,
for disc models A, B and C.  Note that density refers to the surface
number density.}
\label{tab:ranges}
\begin{tabular}{lll}
\hline
Parameter                                &   Used in    &  Range   \\  
\hline
Mass accretion rate $\dot{M}$            &    B, C      &  $10^{-7.5} < \dot{M} < 10^{-3.5}$\,\msunyr \\
Inclination $i$                          &     A, B, C  &  $0 < i < 90\,\degr$ \\
Intrinsic linewidth $\Delta \nu$         & A, B, C      &  $1 < \Delta \nu < 30$\,\kms \\ 
Inner disc radius $R_{\mathrm{i}}$       &      A      &  $1 < R_{\mathrm{i}} < 100$\,R$_{\star}$ \\ 
Inner disc temperature $T_{\mathrm{i}}$  & A      &  $1000 < T_{\mathrm{i}} < 5000$\,K \\
Inner disc density $N_{\mathrm{i}}$      & A      &  $10^{12} < N_{\mathrm{i}} < 10^{25}$\,cm$^{-2}$ \\ 
Temperature exponent $p$                 &     A      &  $-5 < p < 0$ \\
Density exponent $q$                     &   A      &  $-5 < q < 0$  \\ 
\hline
\end{tabular}
\end{center}
\end{table}
\end{center}

\subsection{The fitting procedure}
\label{sec:fitting_procedure}

The best fitting model is determined using the downhill simplex
algorithm, implemented by the \textsc{amoeba} routine of the IDL
distribution.  The input spectra are first continuum subtracted (which
is assumed to be linear given the small range in wavelength), and then
normalised to the peak of the bandhead emission.  Model fits are
compared to the data using the reduced chi-squared statistic,
$\chi_{r}^{2}$, and the error in the data is taken to be the standard
deviation of the flux in the pre-bandhead portion of the spectra.

\smallskip

The fitting routine is repeated with six starting positions spread
across the parameter space to avoid recovering only local minima in
$\chi^{2}_{\mathrm{r}}$.  Table \ref{tab:ranges} shows the ranges over
which we search in parameter space.  The stellar mass, radius and
effective temperature are fixed parameters, and are calculated from
the bolometric luminosity of each object using interpolation of the
main sequence relationships in \citet{martins_2005}, unless otherwise
stated in Table \ref{tab:obs}.  The free parameters of the fit are,
for model A: the inner disc radius, temperature and surface density
$R_{\mathrm{i}}, T_{\mathrm{i}}, N_{\mathrm{i}}$, the temperature and
density exponents $p,q$, the disc inclination $i$ and the intrinsic
linewidth $\Delta \nu$.  For models B and C, the free parameters are:
the mass accretion rate $\dot{M}$, disc inclination $i$ and intrinsic
linewidth $\Delta \nu$.

\subsection{A test of the disc models - W33A}

\begin{figure}

\includegraphics[width=1.0\columnwidth, angle=0,trim=0cm 2.2cm 0cm 1.3cm, clip=true]{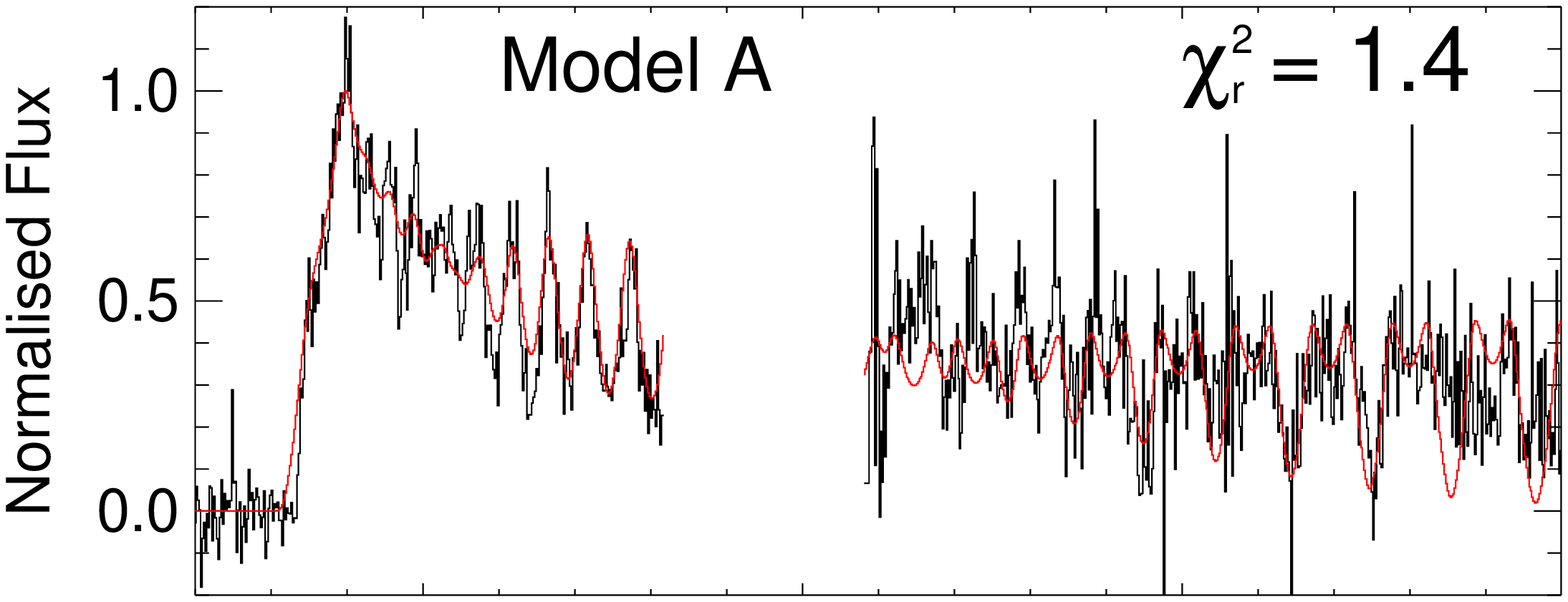}
\includegraphics[width=1.0\columnwidth, angle=0,trim=0cm 2.2cm 0cm 1.3cm, clip=true]{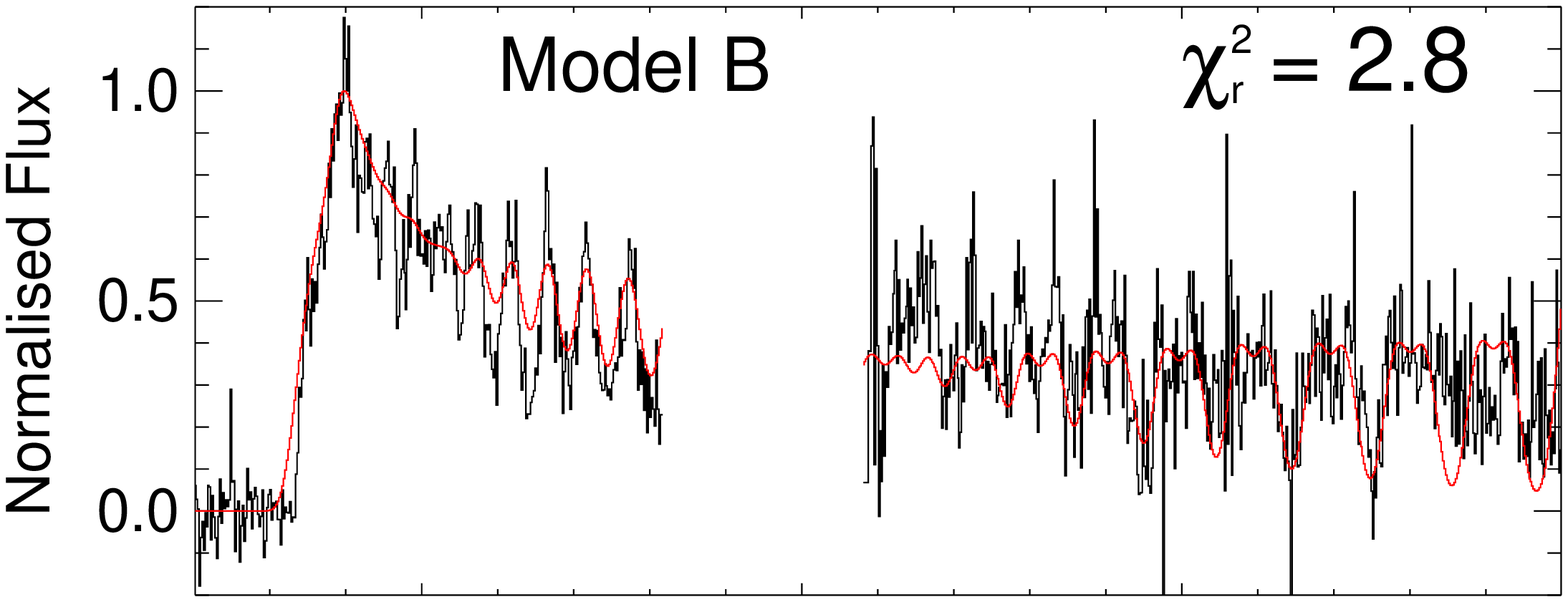}
\includegraphics[width=1.0\columnwidth, angle=0,trim=0cm 0cm 0cm 1.3cm, clip=true]{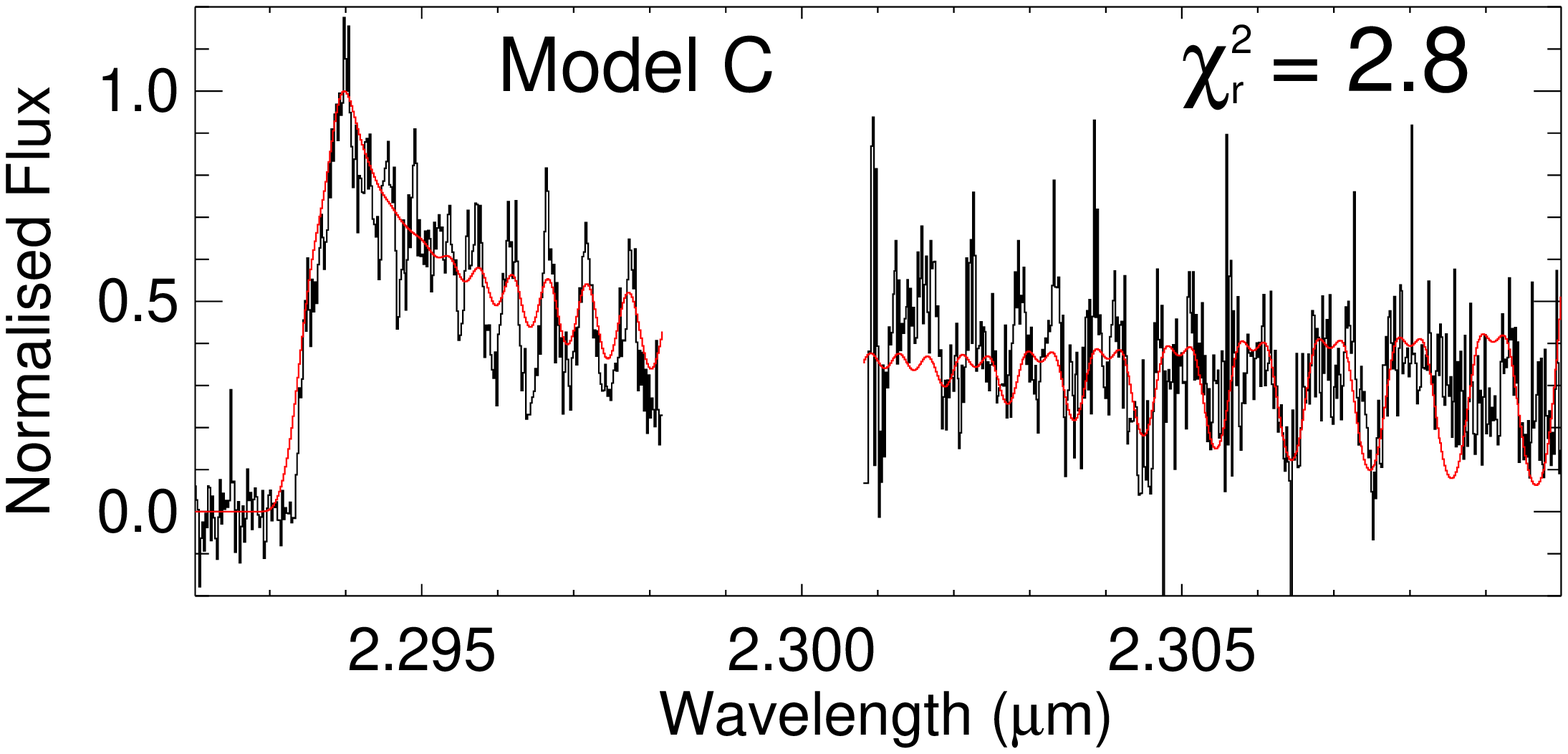}

\caption{Best fitting models to G012.9090$-$00.2607 (W33A) using each disc model.  Best
fitting parameters are shown in Table \ref{tab:discfits}.  Model A
provides the best fit as can be seen from the reduced chi squared
statistic.  Models B and C struggle to reproduce the peak to trough
variation seen in the spectrum between $2.296-2.298\,\micron$.  All
models struggle to fit the initial data points on chip four.}
\label{fig:w33a_all_models}
\end{figure}

The object G012.9090$-$00.2607 (hereafter W33A) is a well known MYSO
that has been studied extensively in recent years
\citep{deWit_2010,wheelwright_2012_visir}, and thus offers a useful
check of our models.  The work of \citet{deWit_2010} determined the
inclination of the system to be $40\degr < i < 70\degr$ through
modelling of the SED.  We have fitted the spectrum of W33A using the
three disc models described in Section \ref{sec:discmodels}, to test
which is most applicable to the circumstellar environment of MYSOs.
Figure \ref{fig:w33a_all_models} shows the data and the best fits, and
Table \ref{tab:discfits} shows the best fitting parameters for each
model.

\begin{center}
\begin{table}
\begin{center}
\caption{Best fitting parameters using disc models A, B and C for the
spectrum of G012.9090$-$00.2607 (W33A).}
\label{tab:discfits}
\begin{tabular}{lllll}
\hline
  Disc Model    & $\dot{M}$                 & $\Delta v$  & $i$         & $\chi^{2}_{\mathrm{r}}$     \\
                       & (\msunyr)                 & (\kms)      & (\degr) &                    \\ 
\hline
  A           &           -                 &   21     & 37            &     1.4         \\
  B           &   $7.76\times10^{-6}$       &   29     & 13            &     2.8        \\
  C           &   $2.14\times10^{-5}$       &   29     & 27            &     2.8        \\
\hline
\end{tabular}
\end{center}
\end{table}
\end{center}

\smallskip

Model A clearly gives the best fit to these data, as can be seen from
the reduced chi-squared statistic.  All models reproduce the peak of
the bandhead and blue side slope well. However, only model A
accurately reproduces the individual line profiles between
2.296--2.298$\,\micron$ with a sufficiently small intrinsic linewidth.
Models B and C both have similar chi-squared values, twice that of
model A. Furthermore, the inclination of model B is too low to agree
with \citet{deWit_2010}.  Model C is more sensitive to high accretion
rates, which allows higher temperatures in the disc to be reached. As
can be seen, higher mass accretion rates are reported than in model B.
The hotter disc means that the CO emission region is located further
out in the disc, and therefore suffers from less rotational
broadening.  This is why model C reports a higher inclination. However,
this is still too low to be consistent with \citet{deWit_2010}.

\smallskip

Model A provides additional parameters in the temperature and density
exponents, and their respective values at the inner edge of the CO
emission region.  This allows us to effectively change the amount of
material within the disc, and provides a better fit to these data.
While models B \& C are both based on physical descriptions of
accretion discs, they may not contain all of the relevant details. For
example, these models also assume emission from a flat disc, but in
reality different disc geometries may need to be considered, such as
flared discs, or discs with discrete vertical layers.  For this
reason, we chose to adopt disc model A as a basis for the fitting
routine for the other MYSOs, as it allows freedom to account for
different emission geometries and is not reliant upon possibly
inaccurate assumptions regarding the temperature and density structure
of these discs.

\smallskip

Adopting disc model A, fits to all objects were obtained as described
in Section \ref{sec:fitting_procedure}.  The errors were calculated by
holding all but one parameters constant at their best fitting value,
and varying the selected parameter until the difference in
$\chi^{2}_{\mathrm{r}}$ was equal to one.  For some parameters, the
1-sigma error values were beyond the range of allowed parameters for
our fitting routine.  For all objects, we attempted the fitting
procedure using data from both chips three and four of the detector.
If a satisfactory fit was not obtained, we limited the wavelength
range of fitting to only include data from chip three and repeated the
fitting procedure.  Tests with individual objects showed that
extrapolation of a fit using only the wavelength range of chip three,
on to chip four, produced similar results to a fit involving both
chips.  This can be seen in several objects (for instance
G296.2654$-$00.3901 and M8E-IR) where the extrapolated fit across chip
four is consistent with the location of the rotational lines that were
not included in the fitting range due to a poor signal-to-noise at
these wavelengths.

\section{Full sample results}
\label{sec:results}

We are able to fit all objects with spectra that have sufficiently
strong CO emission, found to be above approximately six times the rms
noise in the pre-bandhead section of the spectrum.  We fit 8 MYSOs and
the 2 lower mass YSOs across the full chip three and four wavelength
range, and also use only chip three (or a portion of chip three) to
fit the remaining 9 MYSOs.  Three objects (G293.8947$-$00.7825,
G332.0939$-$00.4206 and G339.6816$-$01.2058) were determined to have
CO emission that was too weak, or a signal-to-noise ratio that was too
low, to provide a reliable fit, but for completeness we include their
spectra in Appendix \ref{sec:others}.  The two reclassified low mass
YSO objects, G332.9457$+$02.3855 and G338.5459$+$02.1175, are excluded
from our analysis here and discussed in Appendix \ref{sec:others}.

\smallskip 

Figure \ref{fig:results1} presents the spectra and best fitting models
using disc model A, to each of the MYSOs with CO emission. Objects
where only chip three has been used for the fitting routine are
indicated with a greyed-out chip four region.  Table \ref{tab:results}
shows the best fitting model parameters for each of the objects, with
associated 1-sigma errors. Errors marked with an asterisk should be
considered lower limits, as the full 1-sigma error value was beyond
our allowed parameter range.

\begin{center}
  \begin{table*}
    \begin{center}
      \begin{minipage}{\textwidth}
	\begin{center}
	  \renewcommand{\footnoterule}

	  \caption{Quantities used for the model fitting, and results.
	  The stellar mass, radius and effective temperature are
	  derived from bolometric luminosity using interpolation from
	  the main sequence relationships described in
	  \citet{martins_2005}, unless otherwise stated.  The outer
	  disc radius is defined at the point in the disc in which the
	  temperature drops below 1000\,K, so no error is presented.
	  If an error value is marked with an asterisk, the change in
	  the value of the reduced chi-squared statistic
	  $\chi^{2}_{r}$ was less than one across the allowed
	  parameter range.}
          \label{tab:results}

          \begin{tabular}{p{25.0mm} p{5mm} p{5mm} p{5mm} p{7mm} p{4mm} p{9mm} p{17mm} p{11mm} p{11mm} p{9mm} p{9mm} p{3mm} }
\hline
   Object                               &$M_{\star}$            &  $R_{\star}$          &  $T_{\mathrm{eff}}$  &   $R_{\mathrm{i}}$                 & $R_{\mathrm{o}}$   & $T_{\mathrm{i}}$                   &  $N_{\mathrm{i}}$                              & $p$                               &  $q$                            & $i$                    & $\Delta \nu$                    &  $\chi^{2}_{r}$    \\
                                        &($\mathrm{M}_{\odot}$) & ($\mathrm{R}_{\odot}$)&    $(\mathrm{K})$    &     (au)                  &    (au)   &   (K)                     &   (\cm2)                         &                                        &                                 & (\degr)                &   (\kms)                        &                    \\
\hline
\textbf{MYSOs}\\
G012.9090$-$00.2607        	        &	21.2	        &            7.0        &	33500          & $2.3^{+0.9\ast}_{-1.0}$    & 76        &  $2300^{+50}_{-480}$      & $2.2^{+40}_{-\ast}\times 10^{21}$     & $-0.25^{+0.25\ast}_{-4.75\ast}$  & $-4.10^{+1.80}_{-0.90\ast}$     & $37^{+16}_{-11}$        & $21.0^{+9.0\ast}_{-16}$        &     1.4           \\
G033.3891$+$00.1989	                &	12.3	        &            4.8        &	26500          & $2.1^{+0.1\ast}_{-1.0}$    & 3200      & $4600^{+400\ast}_{-1800}$ & $0.4^{+7}_{-\ast} \times 10^{21}$     & $-0.06^{+0.06\ast}_{-4.94\ast}$  & $-1.70^{+0.30}_{-3.30\ast}$     & $40^{+50\ast}_{-25}$    & $29.0^{+1.0\ast}_{-17}$        &     2.8           \\
G035.1979$-$00.7427	                &	17.7	        &            6.1        &	31200	       & $2.3^{+0.5\ast}_{-2.0}$    & 440       & $3100^{+10}_{-3100\ast}$  & $1.4^{+\ast}_{-\ast}\times 10^{23}$   & $-0.20^{+0.20\ast}_{-4.80\ast}$  & $-4.00^{+1.40}_{-1.00\ast}$     & $36^{+54\ast}_{-20}$    & $2.9^{+10}_{-2.9\ast}$         &     1.8           \\
G270.8247$-$01.1112	                &	12.3	        &            4.8        &	26500          & $1.1^{+0.5\ast}_{-1.0}$    & 93        & $3600^{+10}_{-960}$       & $0.4^{+1.9}_{-0.4} \times 10^{22}$    & $-0.32^{+0.10}_{-0.20}$          & $-1.60^{+0.40}_{-0.70}$         & $89^{+1\ast}_{-40}$     & $18.9^{+21.1\ast}_{-13}$       &     2.8           \\
G282.2988$-$00.7769	                &	11.8	        &            4.7        &	26100          & $1.7^{+0.5\ast}_{-0.6}$    & 9         & $4800^{+10}_{-1500}$      & $0.1^{+2.3}_{-\ast} \times 10^{21}$   & $-0.97^{+0.60}_{-2.20}$          & $-1.40^{+1.40\ast}_{-3.60\ast}$ & $80^{+10\ast}_{-30}$    & $16.3^{+13.7\ast}_{-16.3\ast}$ &     2.8           \\
G287.3716$+$00.6444	                &	17.1	        &            6.0        &	30800	       & $0.1^{+0.1}_{-0.1}$        & 3         & $4200^{+80}_{-720}$       & $9.1^{+\ast}_{-4.2}\times 10^{24}$    & $-0.45^{+0.10}_{-0.10}$          & $-0.88^{+0.88\ast}_{-0.10}$     & $17^{+4}_{-2}$          & $3.1^{+1.2}_{-0.6}$            &     7.1           \\
G296.2654$-$00.3901	                &	9.6	        &            4.0        &	23600          & $1.7^{+0.2\ast}_{-1.0}$    & 530       & $3400^{+60}_{-1280}$      & $0.1^{+3.1}_{-0.0}\times 10^{23}$     & $-0.20^{+0.20\ast}_{-0.40}$      & $-2.20^{+0.90}_{-0.80}$         & $23^{+67\ast}_{-23\ast}$& $28.0^{+2.0\ast}_{-22}$        &     1.3           \\
G305.2017$+$00.2072	                &	20.4	        &            6.8        &	33100	       & $0.6^{+0.3}_{-0.3}$        & 52        & $2930^{+20}_{480}$        & $2.1^{+1.6}_{-1.7} \times 10^{21}$    & $-0.27^{+0.10}_{-0.10}$          & $-1.61^{0.20}_{-0.40}$          & $43^{+13}_{-8}$         & $14.5^{+15.5\ast}_{-8}$        &     4.7           \\   
G308.9176$+$00.1231	                &	31.7	        &            9.0        &	37400          & $0.7^{+1.6}_{-0.6}$        & 9         & $4430^{+30}_{-3200}$      & $4.8^{+\ast}_{-\ast} \times 10^{21}$  & $-0.59^{+0.20}_{-4.80\ast}$      & $-0.14^{+0.14\ast}_{-3.6\ast}$  & $67^{+23\ast}_{-38}$    & $12.6^{+17.4\ast}_{-12.6\ast}$ &     0.9           \\
G310.0135$+$00.3892                     &	21.8	        &            7.0        &	33800	       & $2.8^{+0.5\ast}_{-2.0}$    & 69        & $3760^{+100}_{-1400}$     &$3.8^{+10^{21}}_{-\ast} \times 10^{12}$& $-0.43^{+0.20}_{-4.57\ast}$      & $-0.50^{+0.50\ast}_{-4.5\ast}$  & $67^{+23\ast}_{-41}$    & $25.5^{+4.5\ast}_{-25.5\ast}$  &     2.0           \\
G332.9868$-$00.4871	                &	16.6	        &            5.9        &	30400	       & $0.5^{+0.8}_{-0.3}$        & 7         & $4400^{+120}_{1600}$      & $0.1^{+1.9}_{-\ast} \times 10^{21}$   & $-0.59^{+0.20}_{-0.4}$           & $-0.01^{+0.01\ast}_{-1.3}$      & $78^{+12\ast}_{-38}$    & $29.7^{+0.3\ast}_{-28}$        &  1.9              \\
G338.9377$-$00.4890	                &	7.4	        &            3.4        &       21400          & $0.3^{+0.3}_{-0.2}$        & 1         & $4900^{+100\ast}_{-2200}$ & $0.4^{+2.1}_{-\ast} \times 10^{20}$   & $-1.22^{+0.90}_{-3.78\ast}$      & $-1.63^{+1.63\ast}_{-3.37\ast}$ & $89^{+1\ast}_{-45}$     & $29.9^{+0.1\ast}_{-29.9\ast}$  & 1.2               \\
G347.0775$-$00.3927	                &	7.0	        &            3.3        &	20900          & $0.4^{+1.1\ast}_{-0.3}$    & 3         & $4600^{+10}_{-2000}$      & $0.3^{+4.7}_{-\ast} \times 10^{21}$   & $-0.79^{+0.60}_{-4.21\ast}$      & $-1.14^{+1.14\ast}_{-3.86\ast}$ & $84^{+6\ast}_{-61}$     & $26.5^{+3.5\ast}_{-26.5\ast}$  &   1.9             \\
IRAS 08576$-$4334	                &	6.1$^{\mathrm{\circ}}$&      3.0        &	19200	       & $0.6^{+0.1}_{-0.1}$        & 7         & $4980^{+20\ast}_{-800}$   & $2.8^{+6.4}_{-6.4}\times 10^{21}$     & $-0.71^{+0.05}_{-0.05}$          & $-0.001^{+0.001\ast}_{-0.60}$   & $65^{+4.1}_{-11}$       & $15.7^{+2.4}_{-3}$             &    2.9            \\
IRAS 16164$-$5046	                &	34.9     	&            9.8        &	38300	       & $1.9^{+0.1}_{-0.1}$        & 5         & $4380^{+10}_{-1900}$      & $0.1^{+2.1}_{-\ast} \times 10^{22}$   & $-1.45^{+0.90}_{-3.55\ast} $     & $-1.37^{+1.37\ast}_{-3.63\ast}$ & $62^{+16}_{-12}$        & $11.9^{+18.1\ast}_{-8}$         &     1.7           \\
IRAS 17441$-$2910                       &	56.9        	&            14.3       &	41300          & $6.0^{+0.6\ast}_{-0.1}$    & 13        & $3880^{+10}_{-400}$       & $3.8^{+0.2}_{-\ast} \times 10^{21}$   & $-1.72^{+1.30}_{-0.10}$          & $-1.00^{+1.00\ast}_{-4.00\ast}$ & $53^{+2}_{-16}$         & $3.7^{+0.6}_{-3.7\ast}$        &     0.6           \\
M8E-IR$^{\mathrm{\ddag}}$                  &       13.5$^{\mathrm{\dag}}$&      5.1        & 	27800	       & $1.5^{+0.9\ast}_{-0.4}$    & 54        & $2300^{+10}_{-320}$       & $0.2^{+10}_{-\ast} \times 10^{20}$    & $-0.24^{+0.24\ast}_{-4.76\ast}$  & $-2.16^{+1.1}_{-2.84\ast}$      & $14^{+10}_{-14\ast}$    & $49^{+11\ast}_{-13}$           &     2.1           \\
\hline
\multicolumn{3}{l}{\textbf{Non-MYSOs} (see Appendix \ref{sec:others})}\\
G332.9457$+$02.3855$^{\mathrm{\S}}$     &	0.5	        &            3.0        &	5000           & $0.7^{+0.7\ast}_{-0.6}$     & 4         & $3240^{+10}_{-880}$       & $2.5^{+5.3}_{-\ast} \times 10^{20}$   & $-0.73^{+0.73\ast}_{-4.27\ast}$  & $-2.60^{+2.60\ast}_{-2.40\ast}$ & $66^{+24\ast}_{-66\ast}$& $29.8^{+0.2\ast}_{-29.8\ast}$  &   2.7             \\
G338.5459$+$02.1175$^{\mathrm{\S}}$     &	0.5	        &            3.0        &	5000           & $1.2^{+0.2\ast}_{-0.4}$     & 61        & $5000^{+0\ast}_{-1000}$   & $1.5^{+6.0}_{-0.9} \times 10^{22}$     & $-0.36^{+0.00}_{-0.20}$          & $-3.77^{+1.40}_{-1.23\ast}$     & $87^{+3\ast}_{-61}$     & $28.4^{+1.6\ast}_{-11}$        & 3.9               \\
\hline

  \end{tabular} 
	\end{center}

 	\small{$\mathrm{\S}$: Originally thought to be an MYSO prior
 	to observations, but subsequent analysis showed that the
 	bolometric luminosity is too low for this to be the case.  The
 	stellar mass, radius and effective temperature of these
 	objects is therefore estimated from typical T Tauri star
 	values \citep{covey_2011}.} \\
	\small{$\mathrm{\dag}$: Taken from \citet{linz_2009}, which is based on the best fitting model from \citet{robitaille_2007}.}\\
        \small{$\mathrm{\circ}$: Determined from the position in the K versus J$-$K diagram of \citet{bik_2006}, as in \citet{wheelwright_2010}.}\\
        \small{$\mathrm{\ast}$: The value of $\chi^{2}_{r}$ did not change by one over the allowed parameter range.}\\
        \small{$\mathrm{\ddag}$: The wavelengths fitted and ranges of allowed fit parameters were altered to obtain a good fit, see text for details.}\\

 \end{minipage}
    \end{center}
  \end{table*}
\end{center}

\begin{figure*}
\centering

\includegraphics[width=\columnwidth]{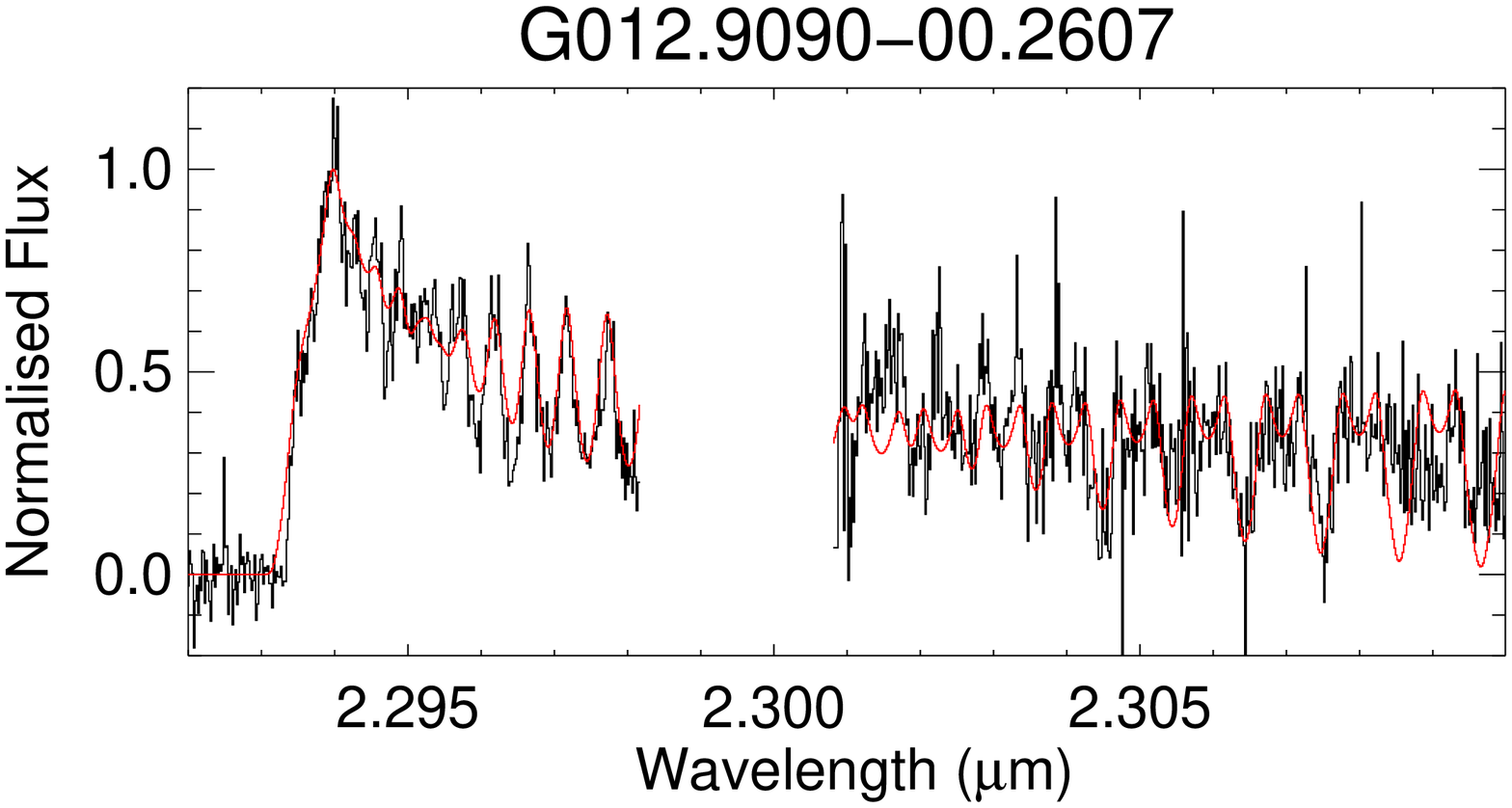}
\includegraphics[width=\columnwidth]{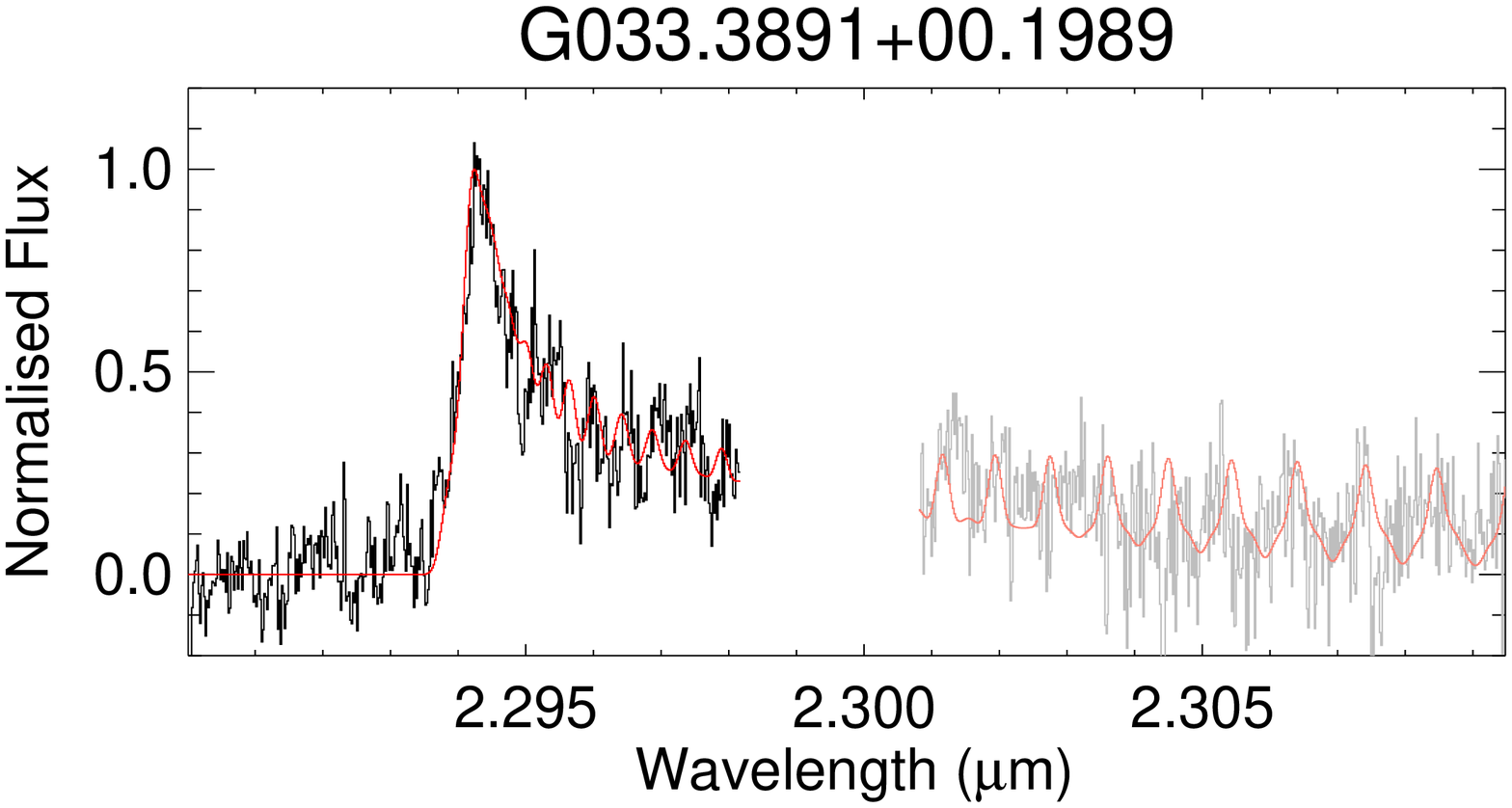}

\includegraphics[width=\columnwidth]{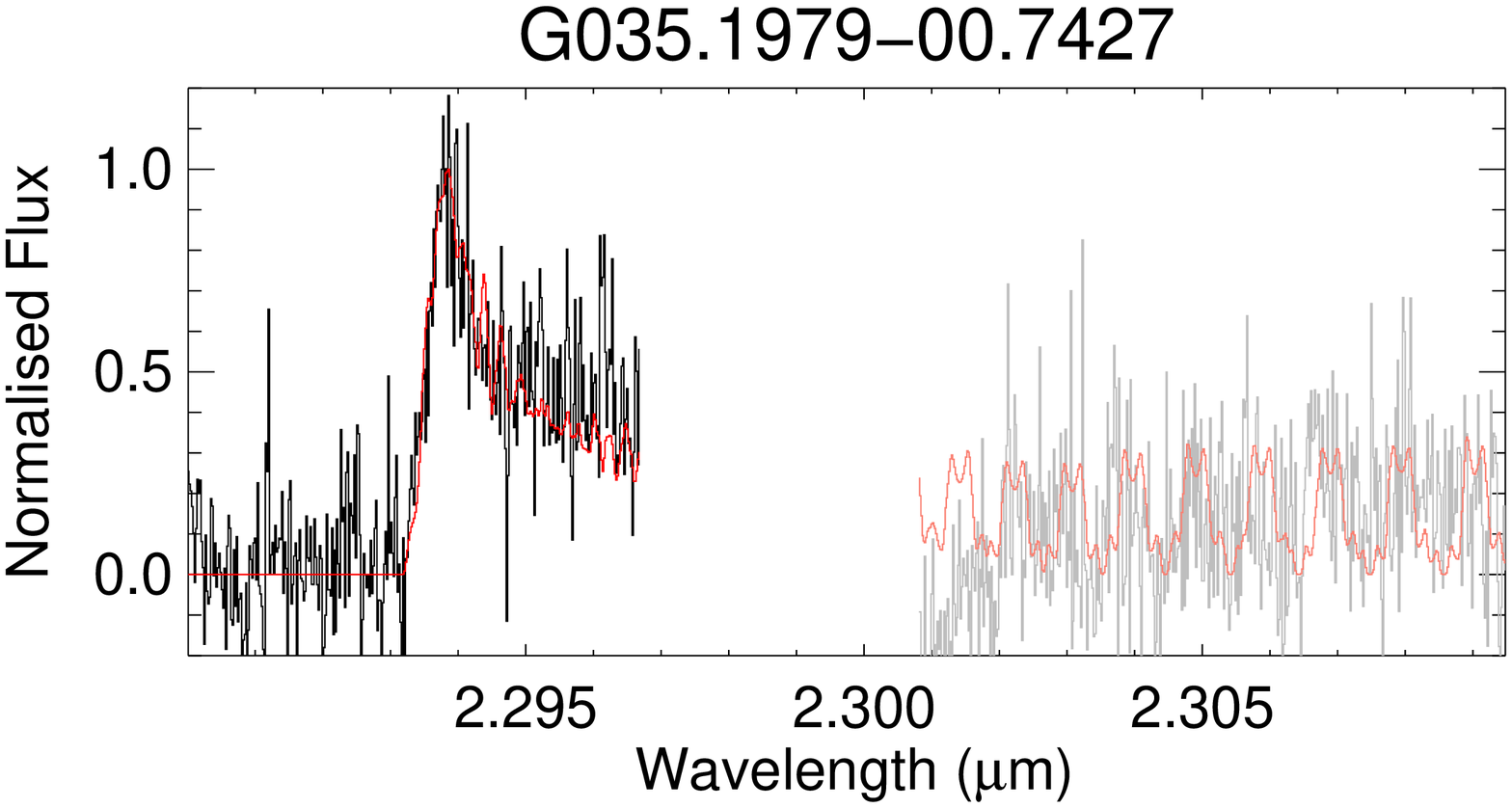}
\includegraphics[width=\columnwidth]{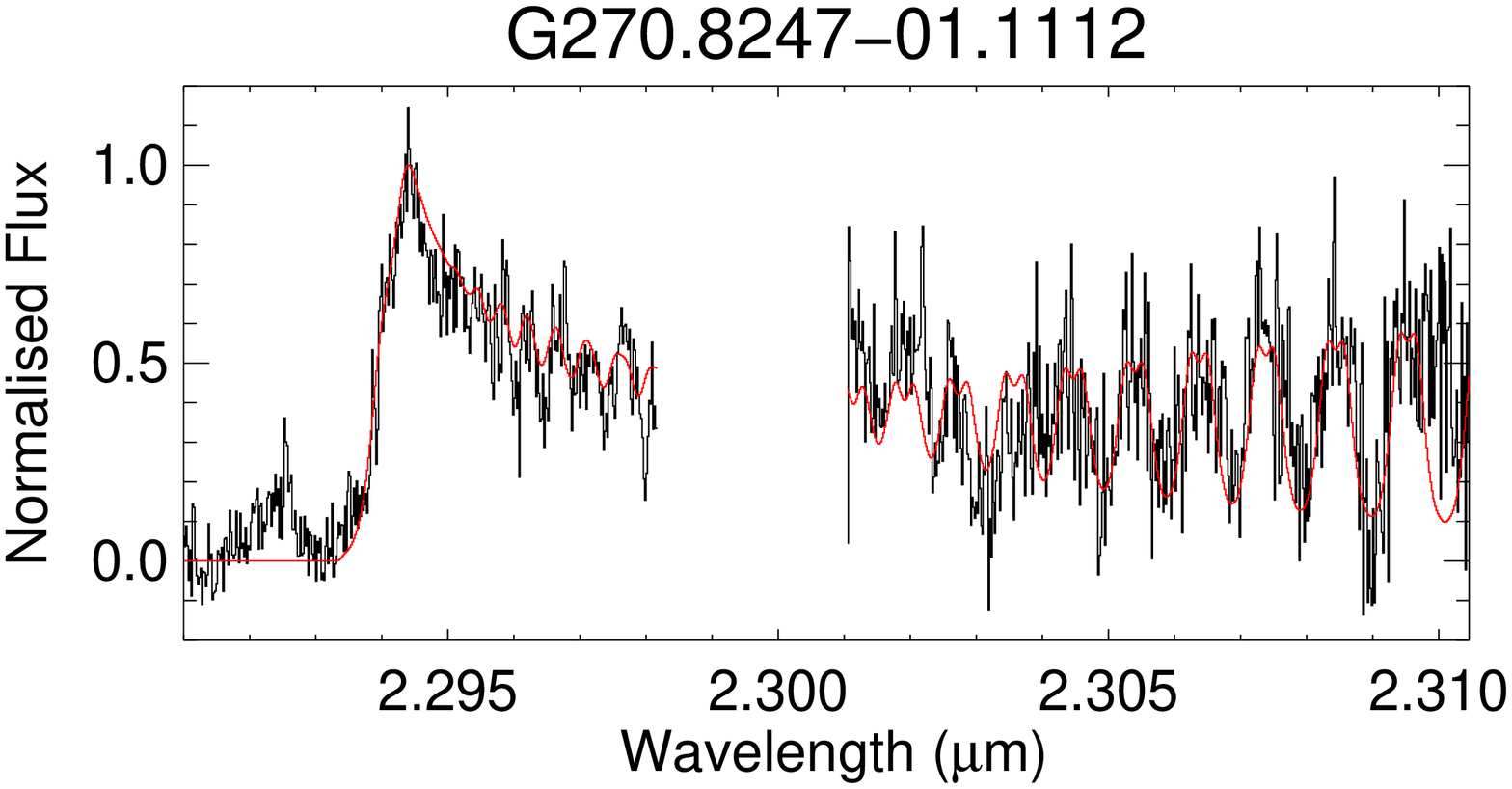}

\includegraphics[width=\columnwidth]{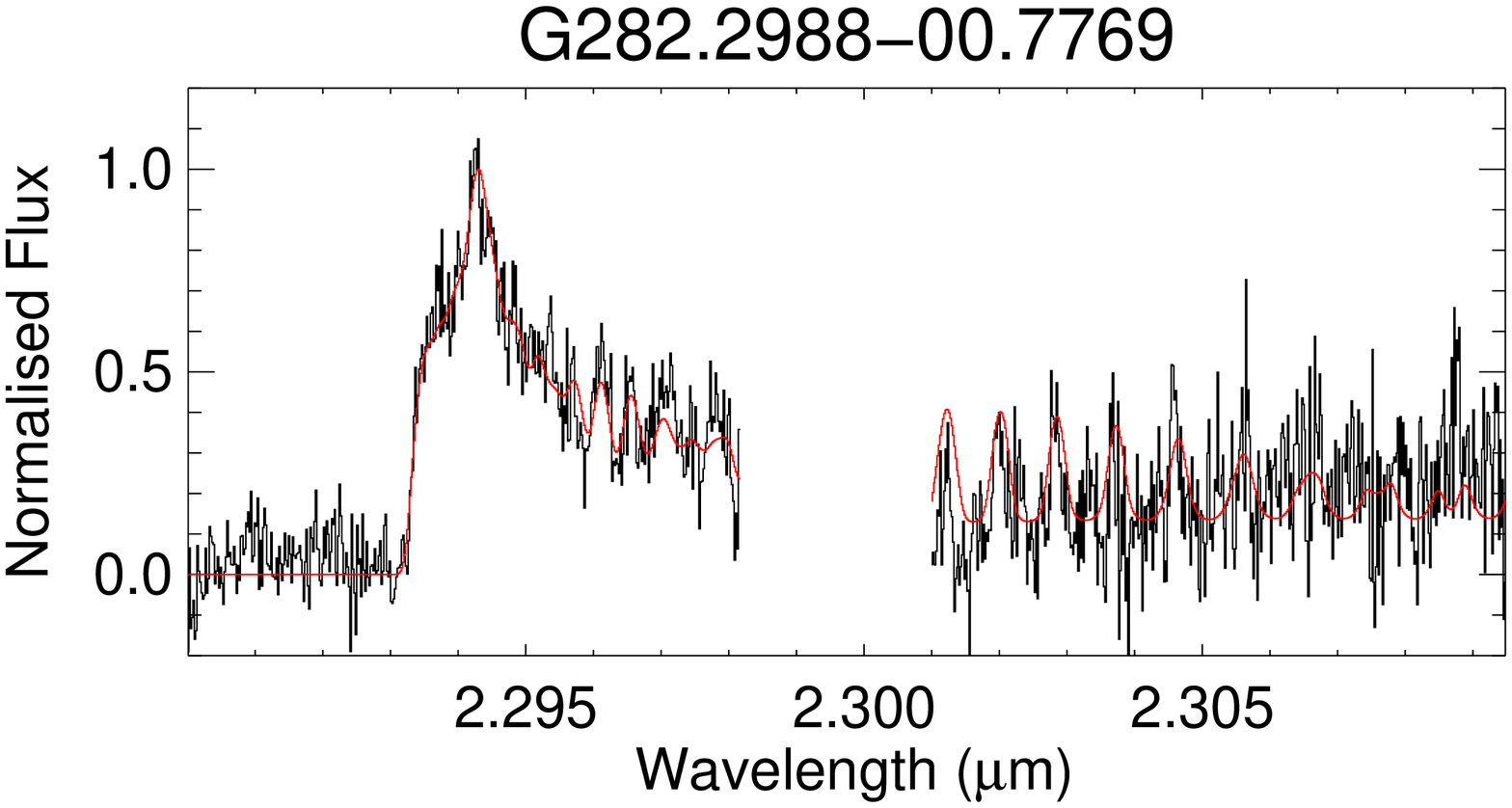}
\includegraphics[width=\columnwidth]{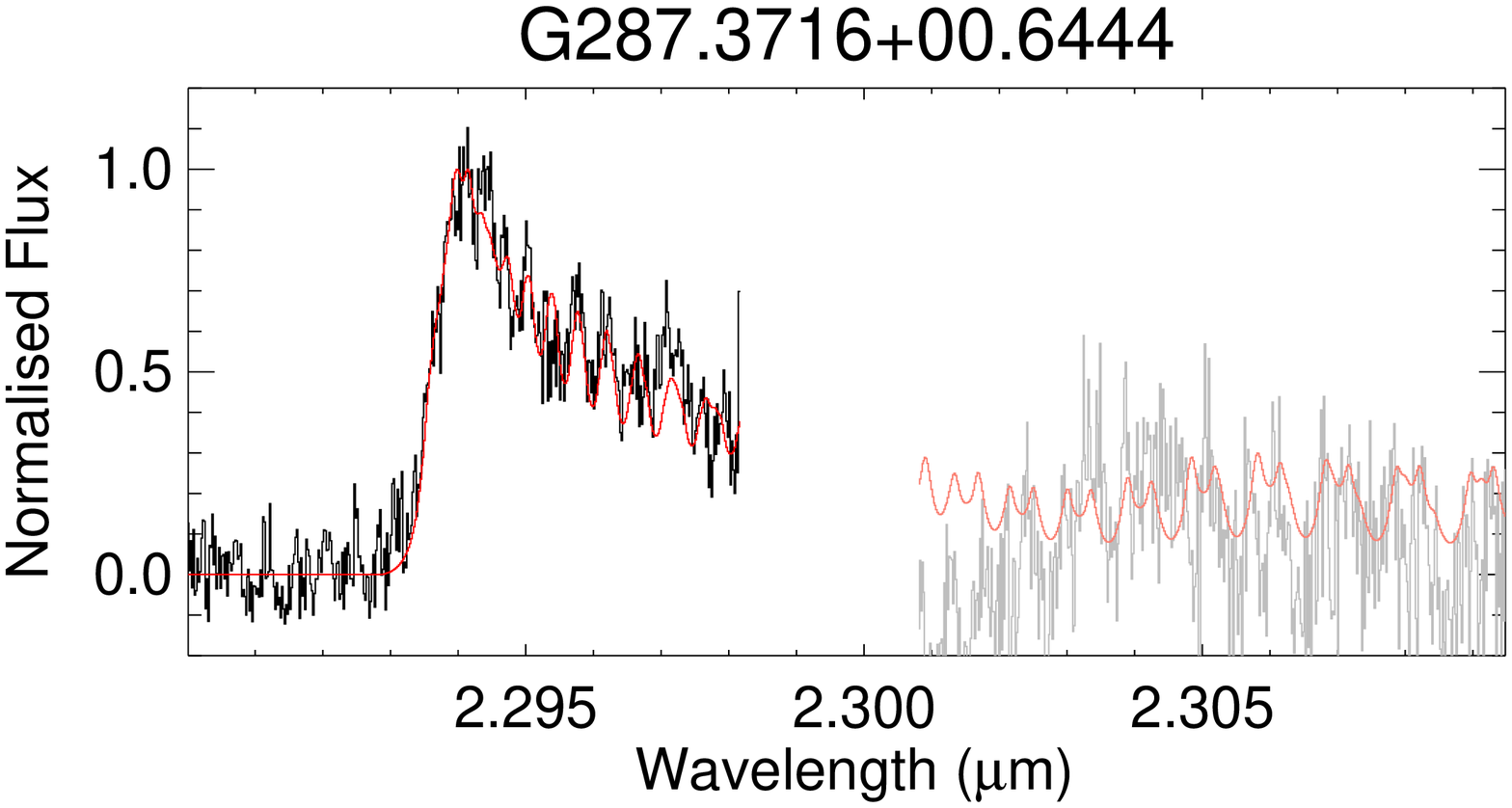}

\includegraphics[width=\columnwidth]{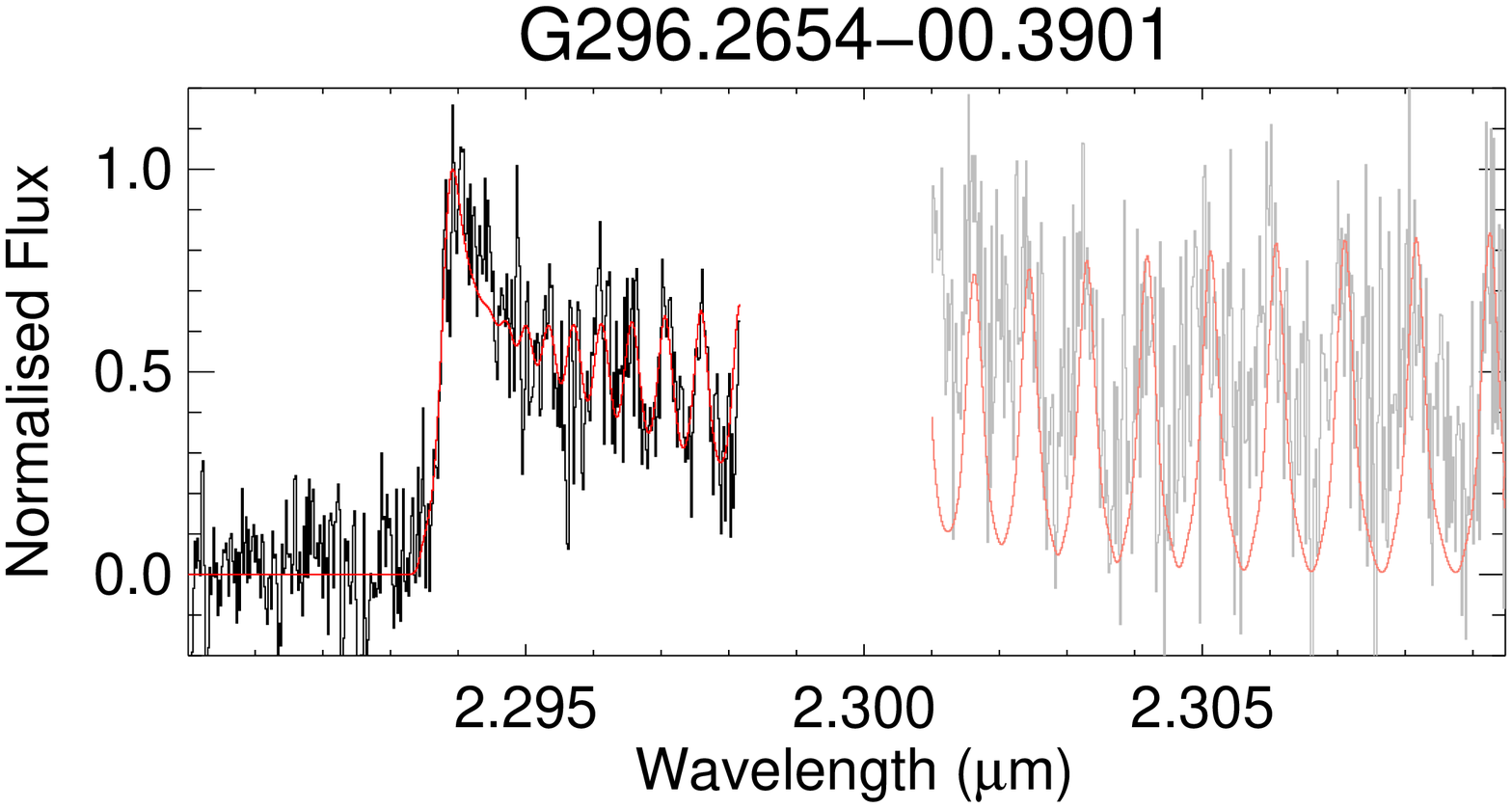}
\includegraphics[width=\columnwidth]{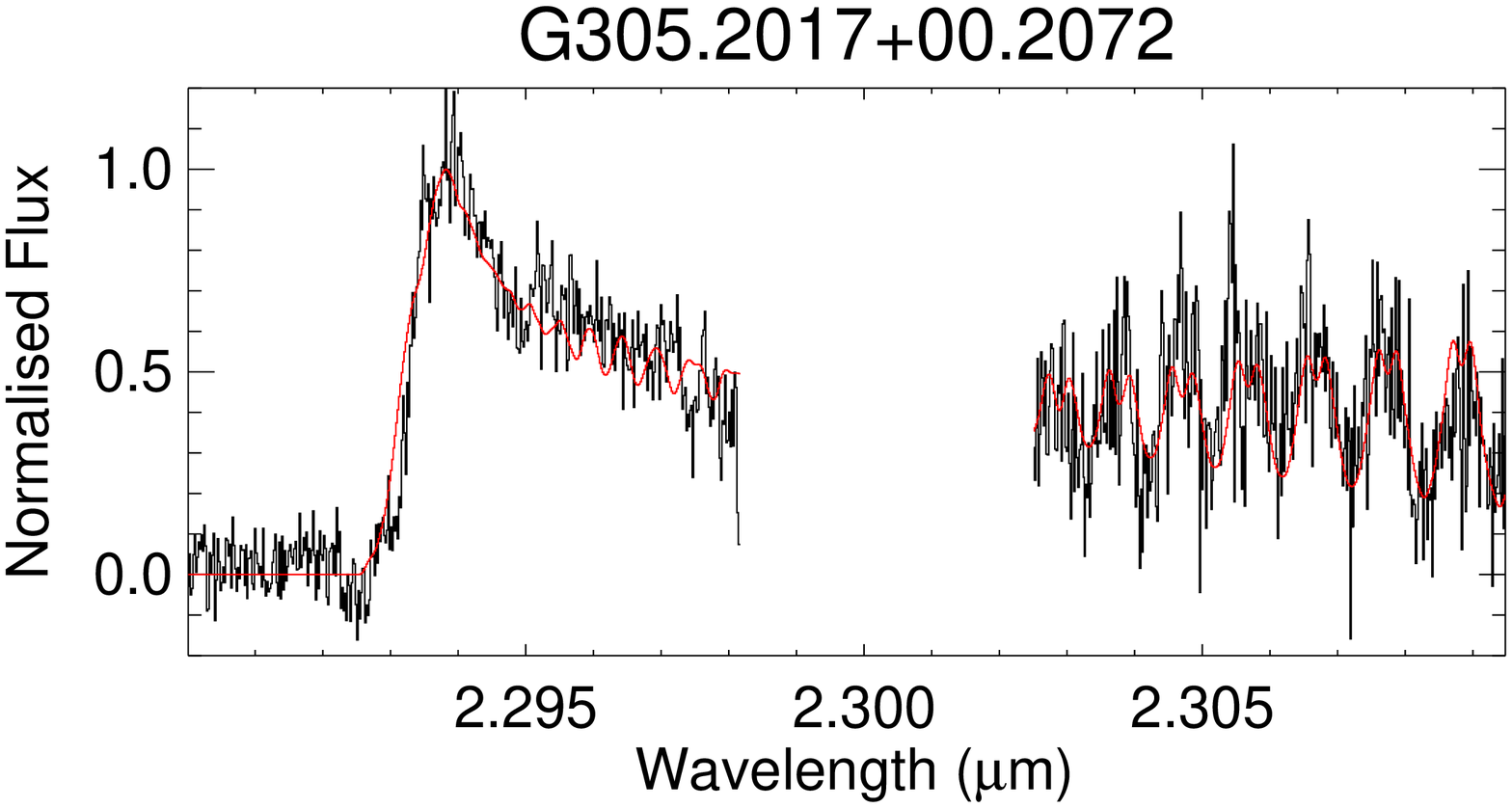}

\includegraphics[width=\columnwidth]{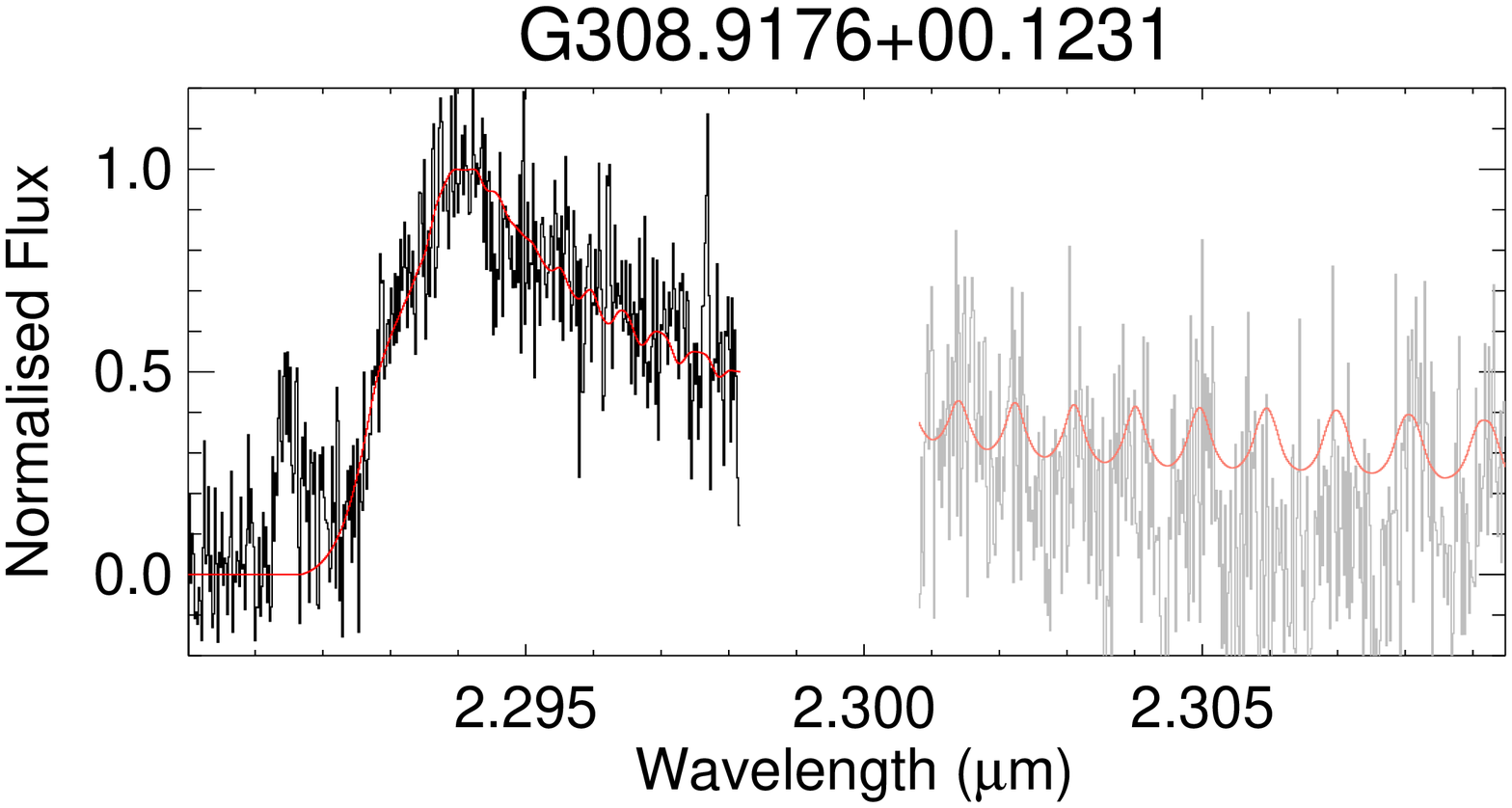}
\includegraphics[width=\columnwidth]{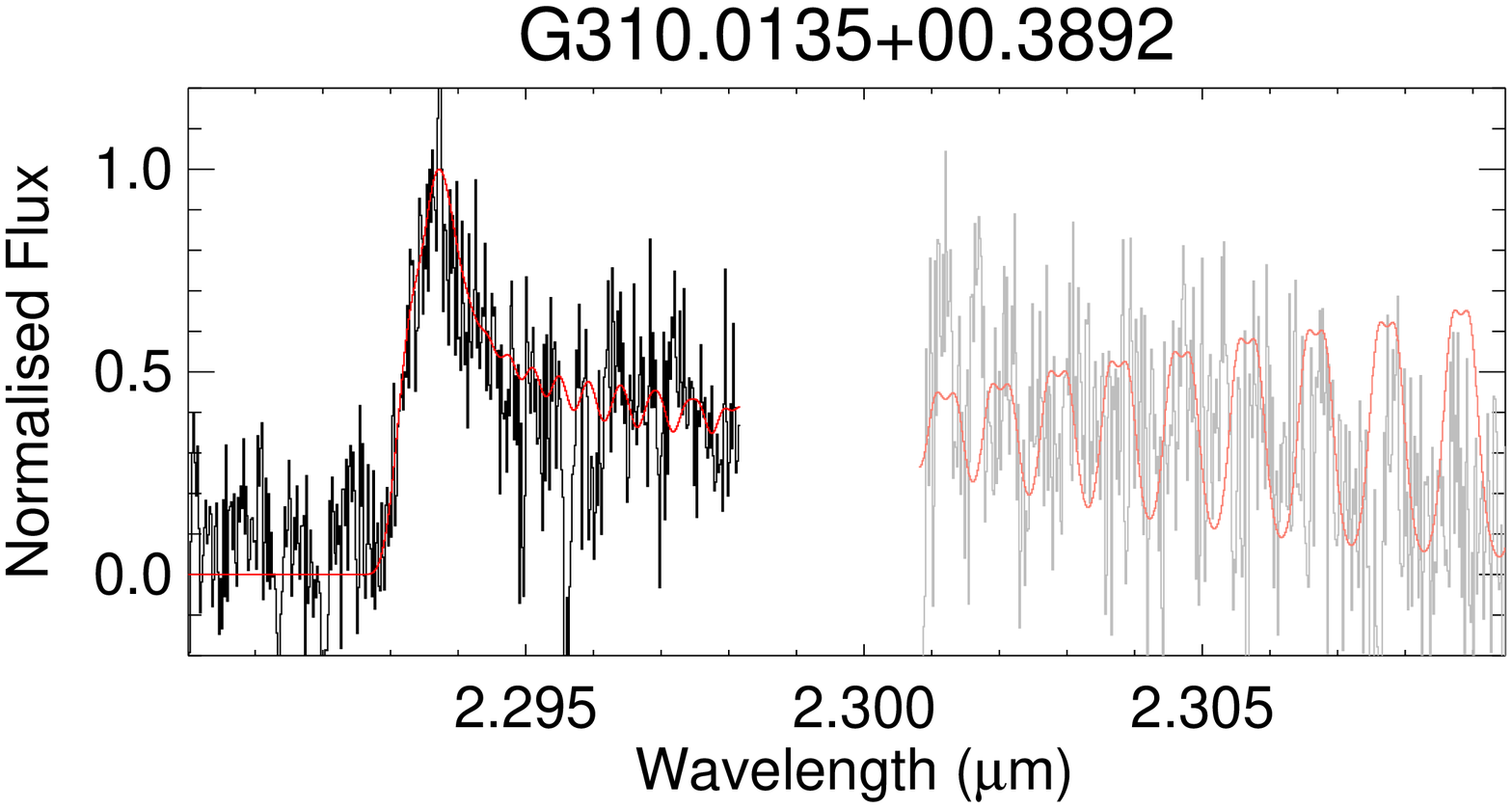}

\caption{Spectra (black) and model fits (red) to the CO emission of
  the objects in the sample, using disc model A. Data that are grayed
  out are not included in the fitting procedure because of poor
  quality, but have been included for completeness.  For
  G035.1979$-$00.7427 and G305.2017$+$00.2072, small ranges of the
  spectrum at the edges of the detectors were removed because of
  excessive noise in the data.}
\label{fig:results1}
\end{figure*}

\begin{figure*}
\centering

\includegraphics[width=\columnwidth]{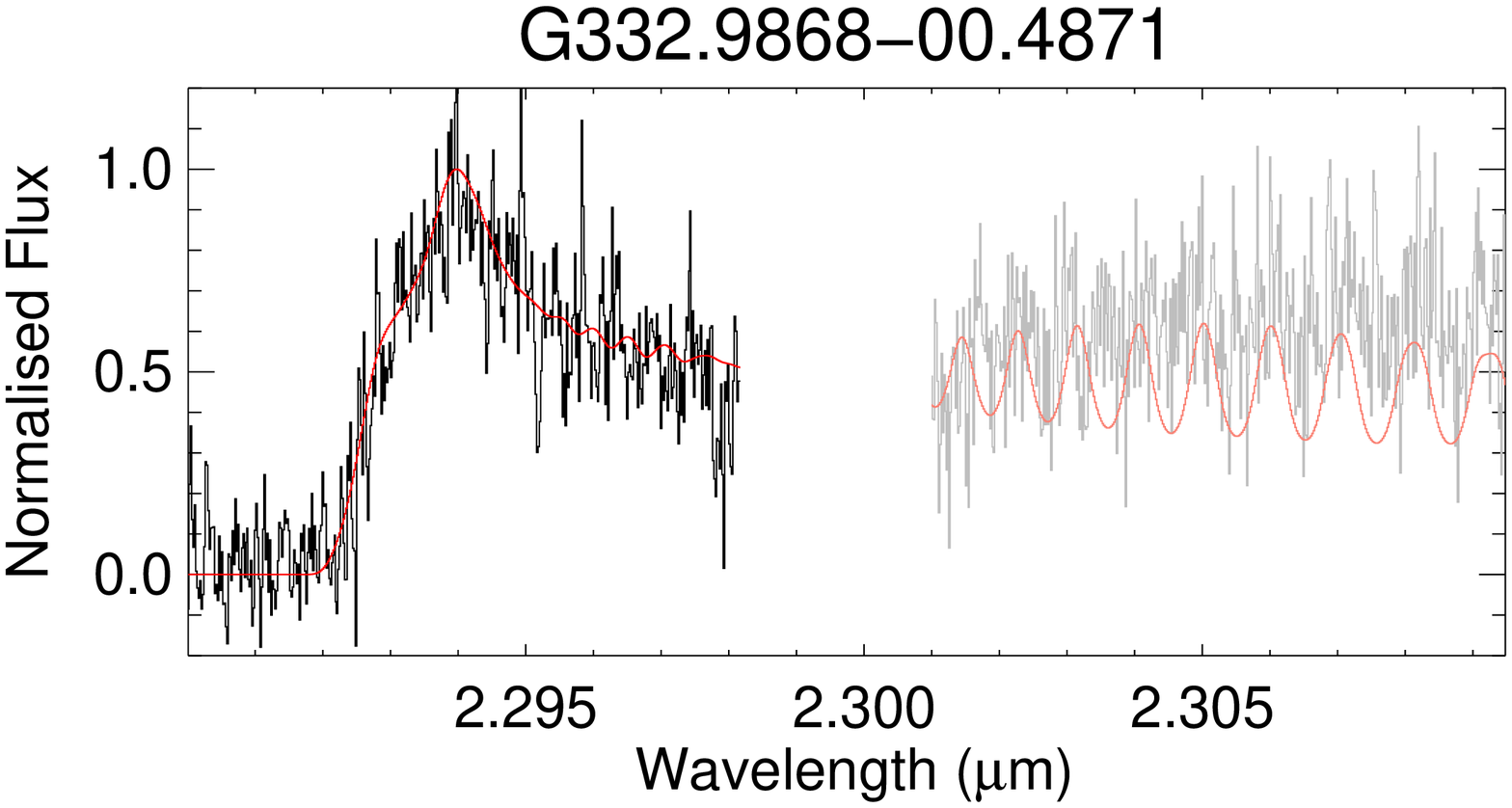}
\includegraphics[width=\columnwidth]{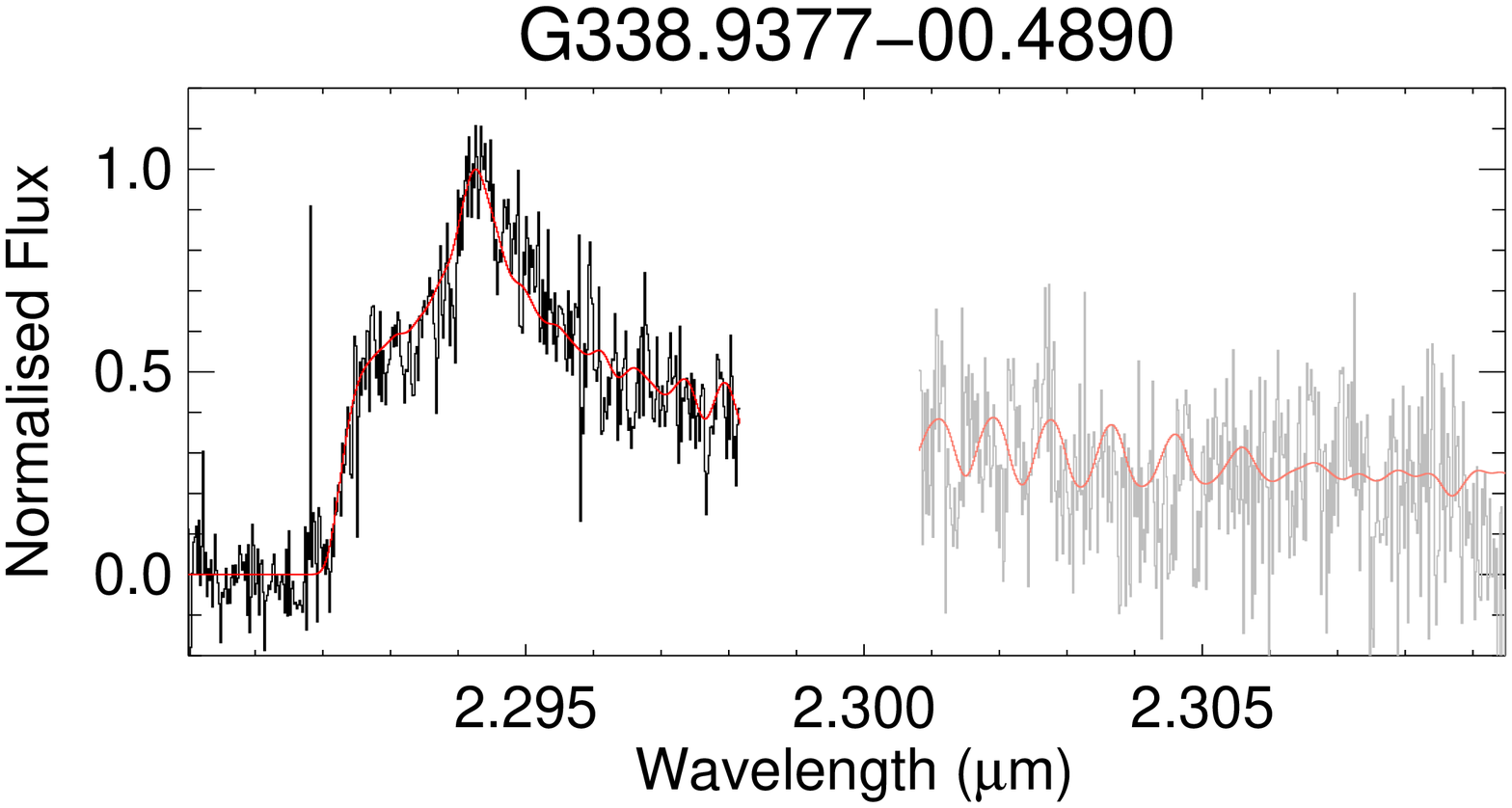}

\includegraphics[width=\columnwidth]{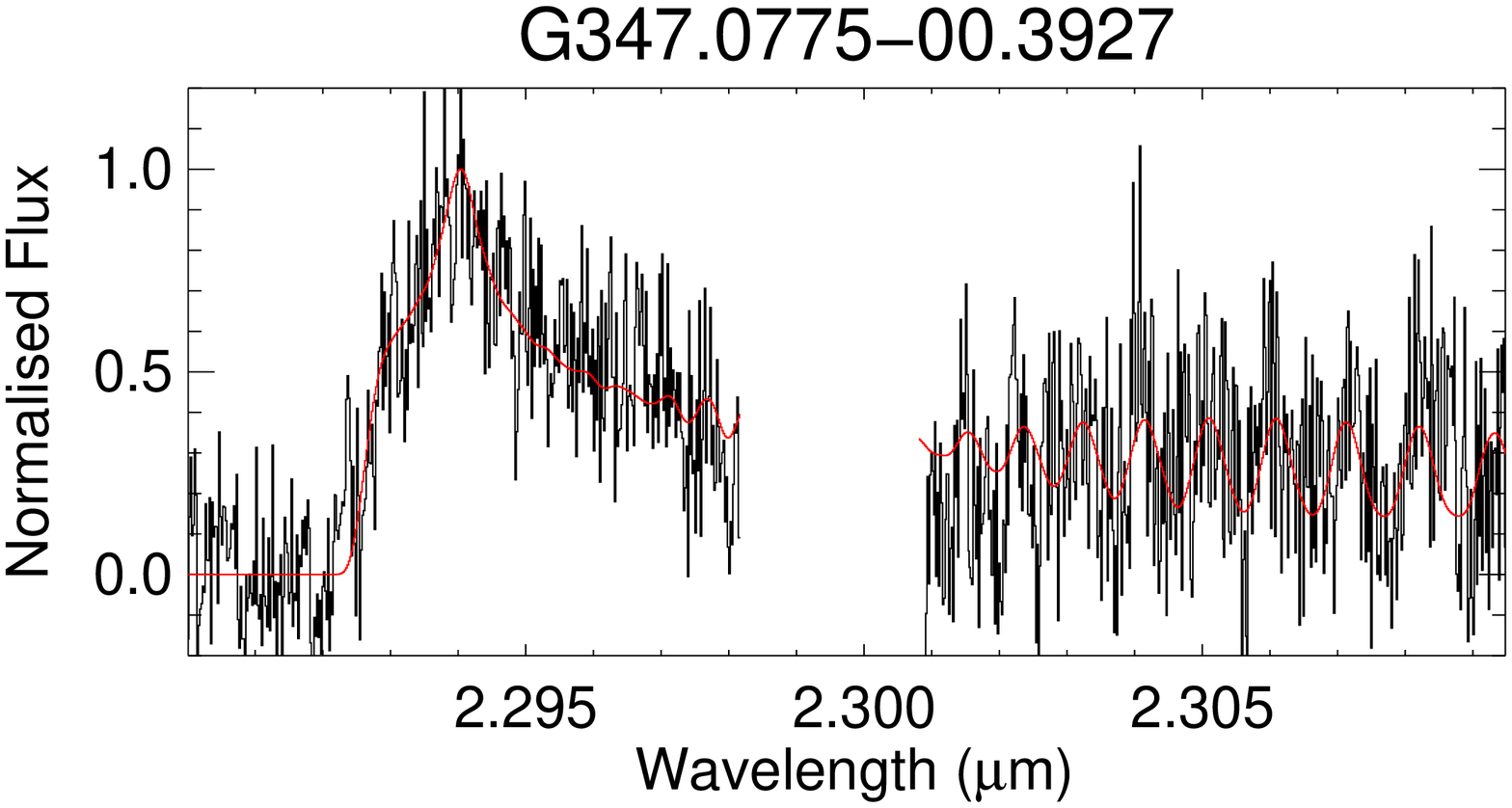}
\includegraphics[width=\columnwidth]{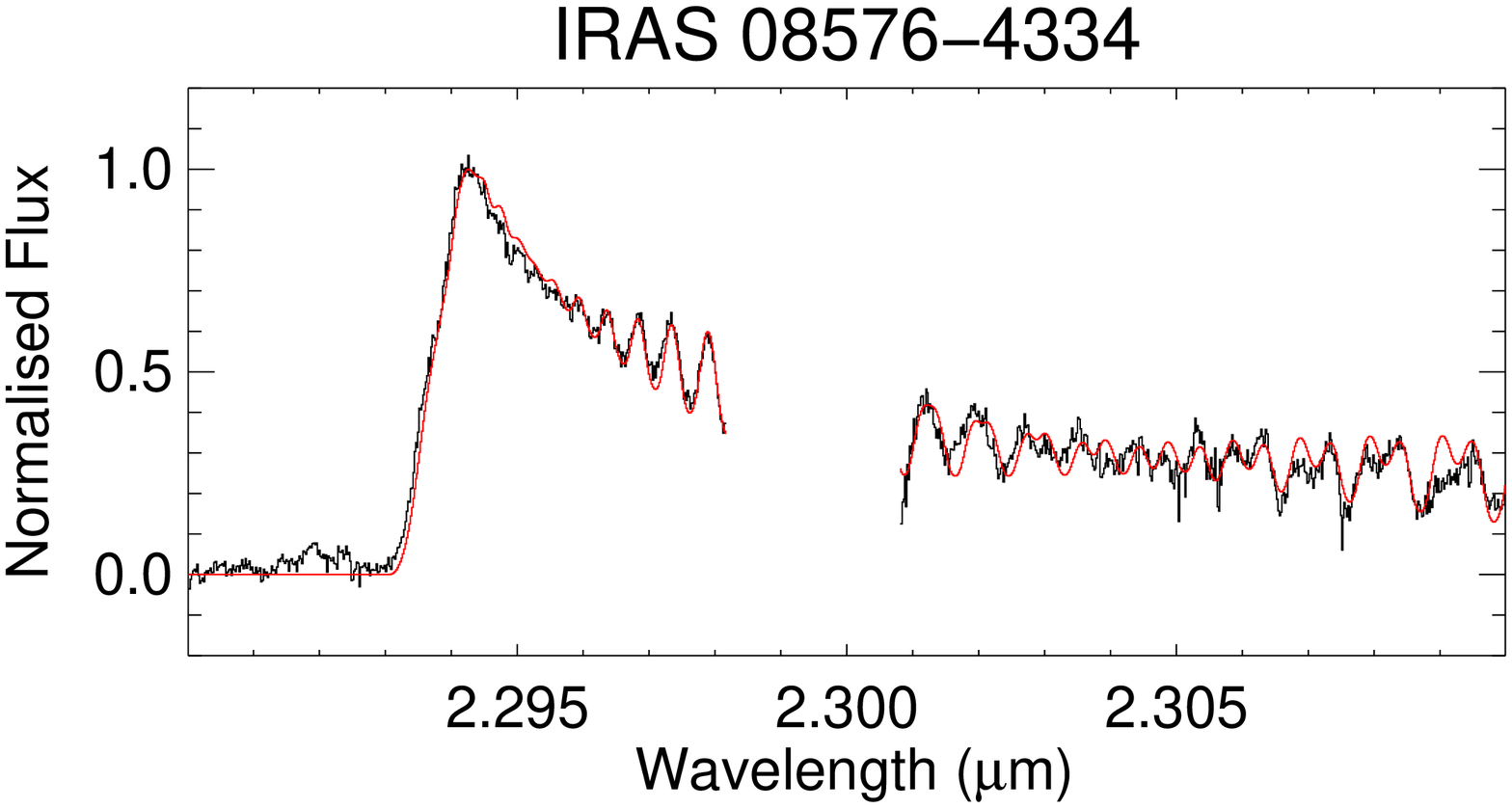}

\includegraphics[width=\columnwidth]{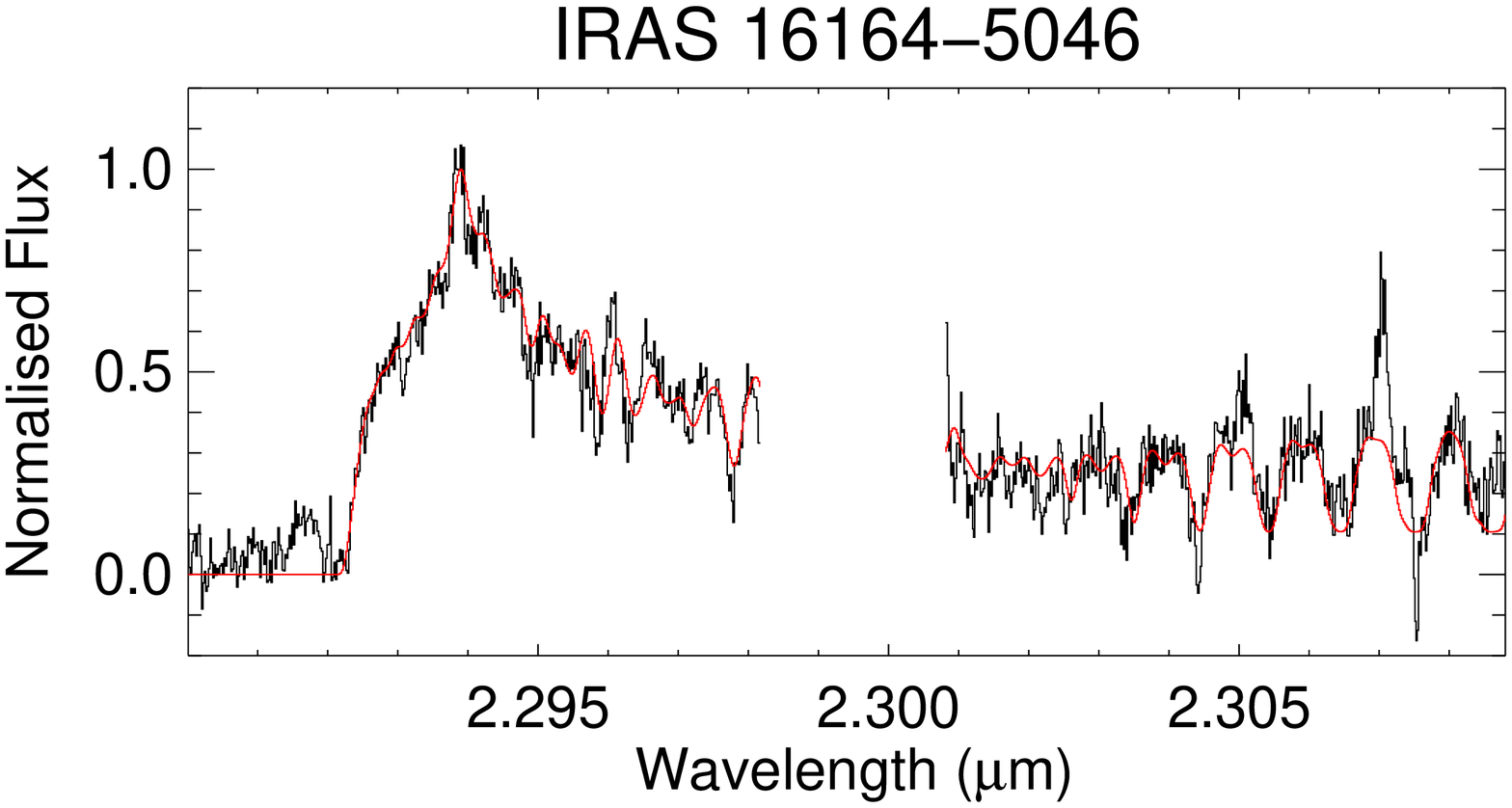}
\includegraphics[width=\columnwidth]{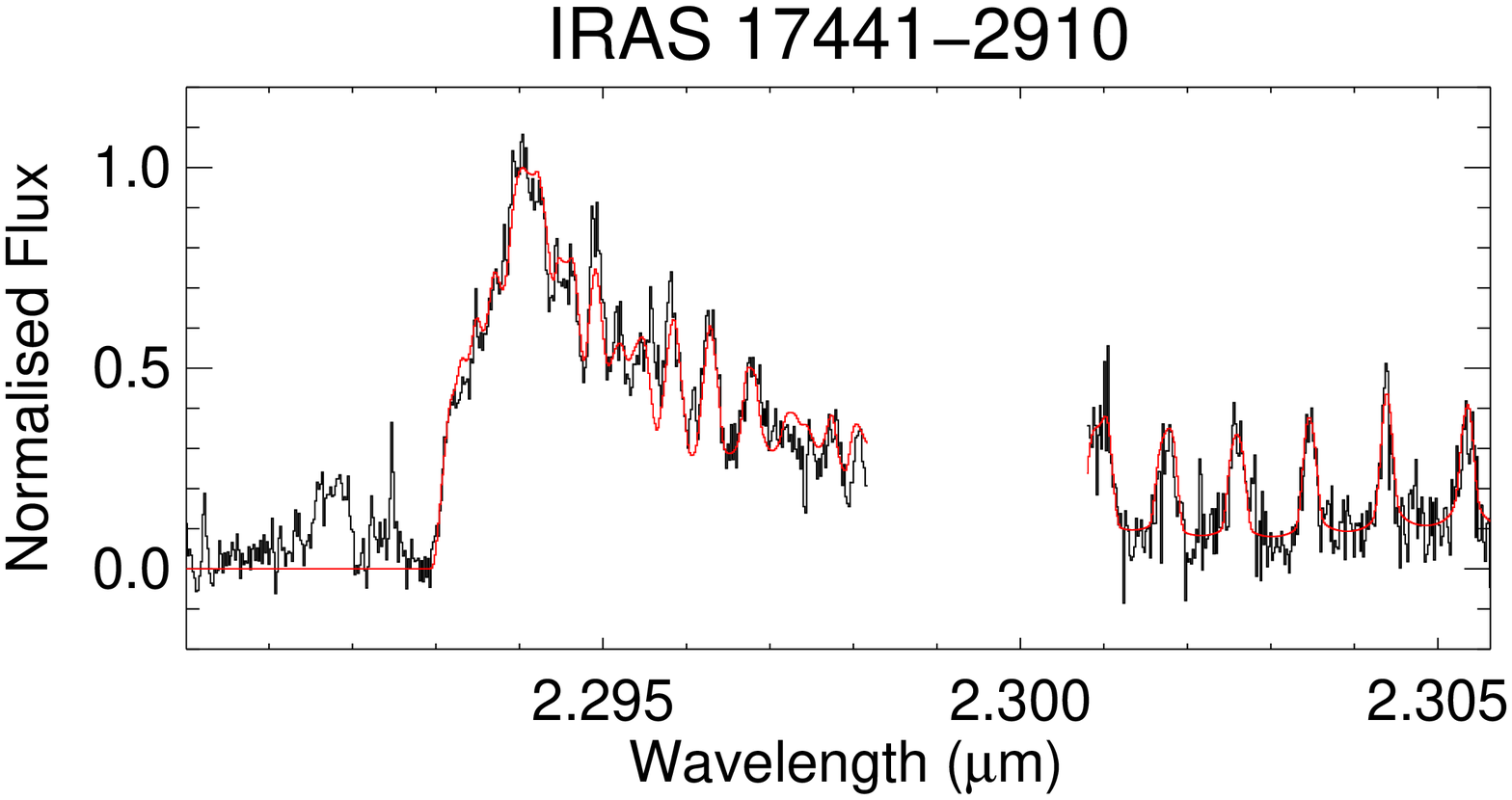}

\includegraphics[width=\columnwidth]{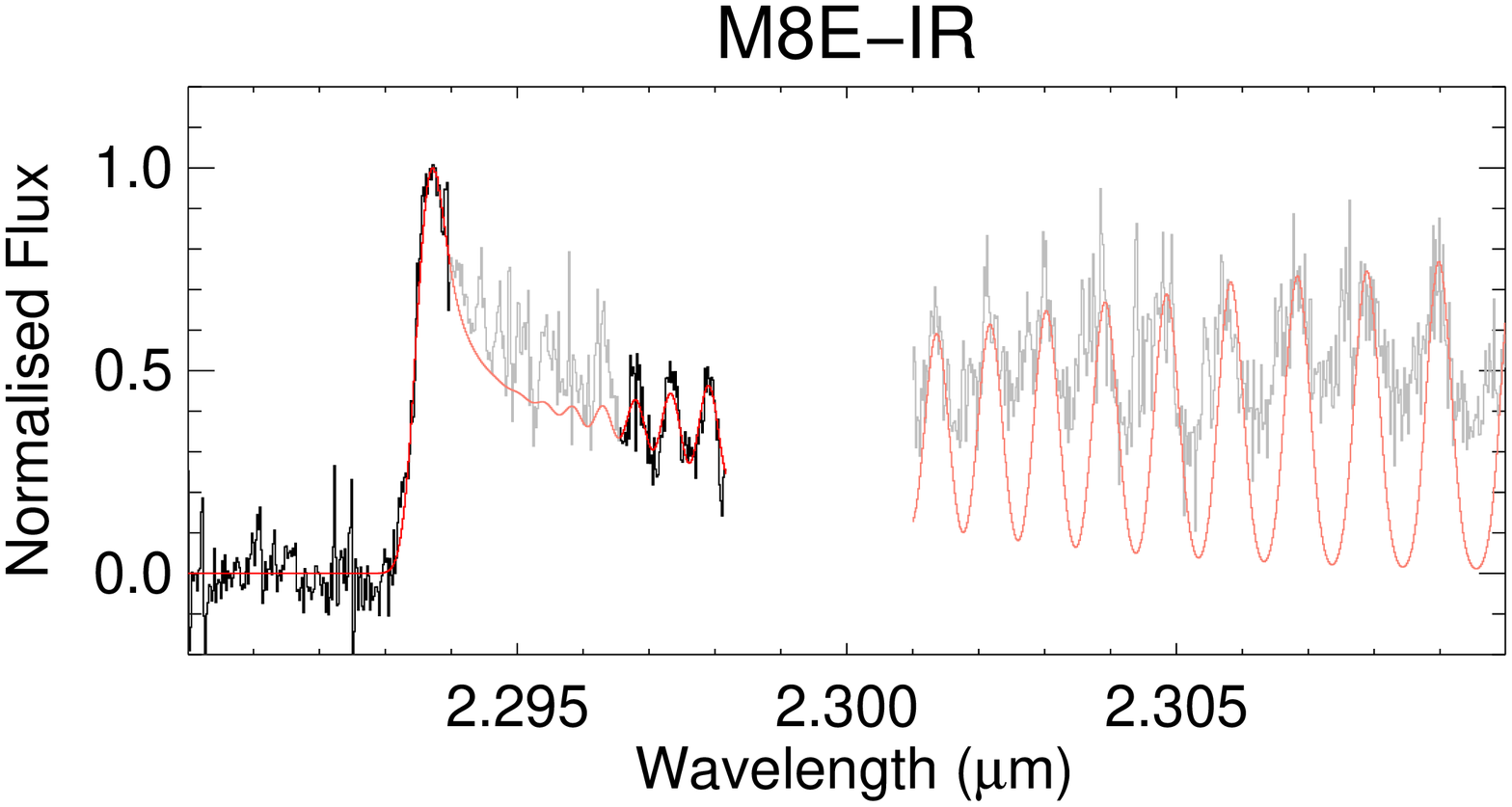}

\contcaption{Spectra (black) and model fits (red) to the CO emission
  of the objects in the sample, using disc model A.  Data that are
  grayed out are not included in the fitting procedure because of poor
  quality, but have been included for completeness.  The range of
  fitting for M8E-IR was restricted to obtain a good fit, see Appendix
  \ref{sec:compare} for full discussion.  For IRAS 17441$-$2910, the
  portion of the spectrum beyond $2.306\,\micron$ was not used because
  an artificial drop in flux was caused by the detector.}
\label{fig:results2}
\end{figure*}

\smallskip
  
In general, across the 7 objects whose inclinations have been
constrained before, our best fitting parameters agree with the
previous results, within error margins.  Comments and comparisons on
an object by object basis are presented in Appendix \ref{sec:compare}.
The majority of objects have discs beginning within a few astronomical
units of the stellar surface, and inner disc temperatures close to the
dissociation temperature of CO (5000\,K). The distribution of the
inner surface densities of the disc has a geometric average of
$\bar{N_{\mathrm{i}}} = 5.5 \pm 8 \times10^{20}\,\mathrm{cm}^{-2}$.
It should be noted that even though CO should be dissociated by
stellar UV flux at these small distances, we find as in
\citet{bik_2004} and \citet{wheelwright_2010} that the density is
sufficient for self-shielding to occur ($N > 10^{15}\mathrm{cm}^{-2}$,
\citealt{vandishoeck_1988}).\\

\smallskip

\begin{figure}
\centering
\includegraphics[width=\columnwidth, angle=0,trim=1cm 1cm 1cm 2cm, clip=true]{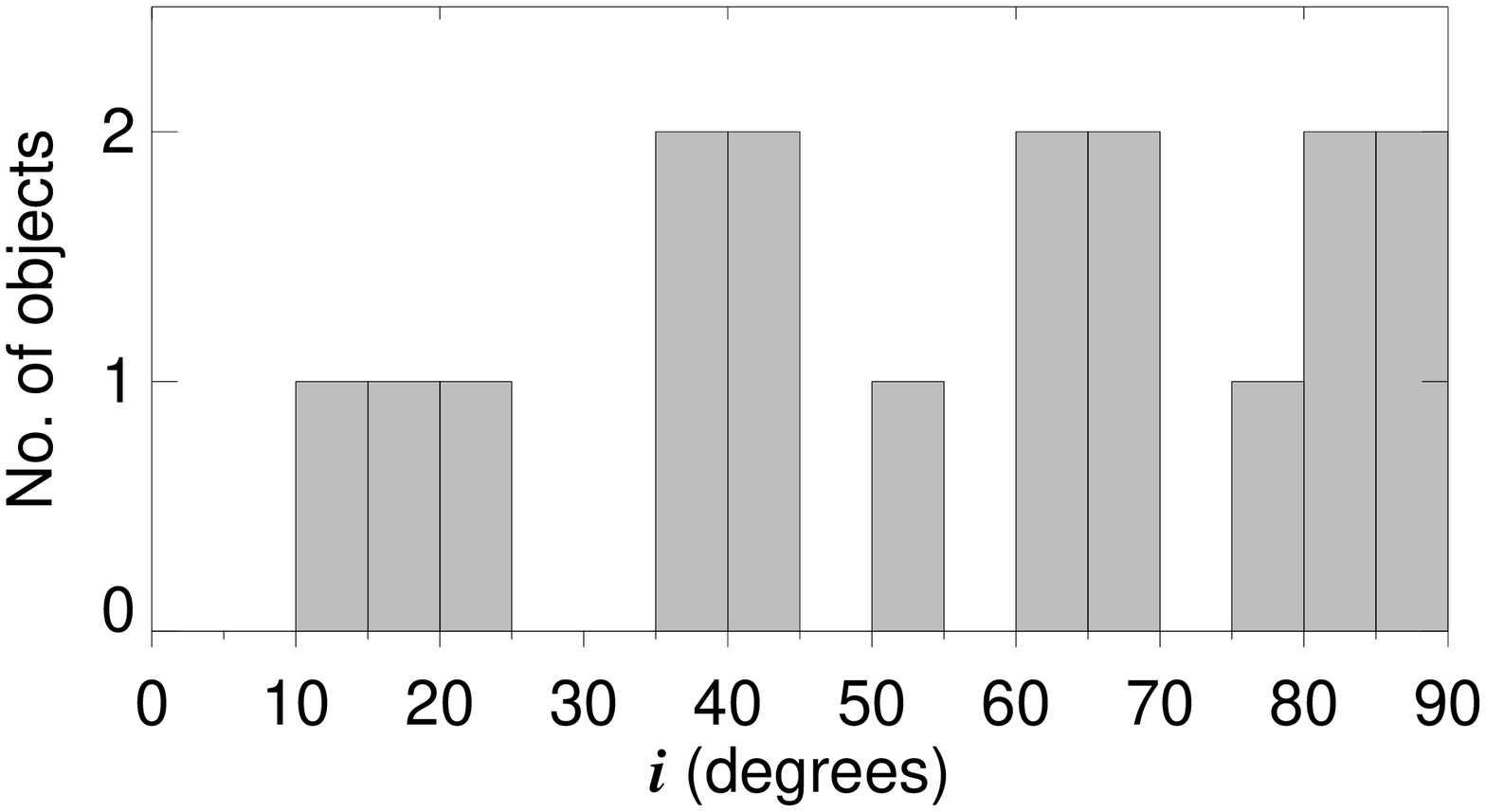}
\includegraphics[width=\columnwidth, angle=0,trim=1cm 1cm 1cm 2cm, clip=true]{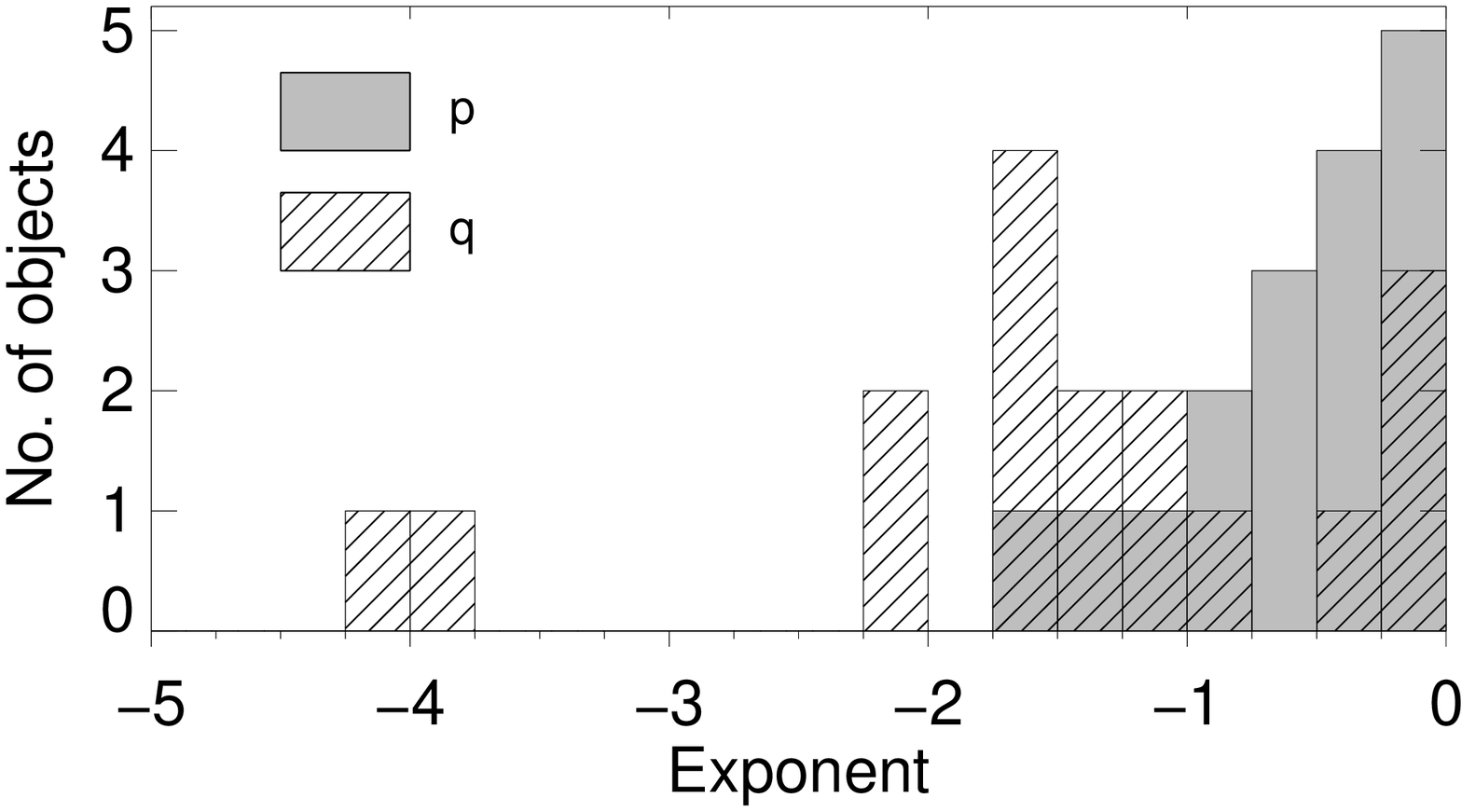}
\caption{Distributions of inclinations (top), and temperature and
surface density exponents ($p$ and $q$, respectively, bottom)
determined from the fits to the 17 MYSOs with detected CO bandhead emission.}
\label{fig:incpq_histo}
\end{figure}

Figure \ref{fig:incpq_histo} shows the distribution of the best
fitting inclinations and the temperature and density exponents of the
MYSOs.  While our inclinations are consistent with most previously
published data for objects, IRAS 08576$-$4334 is not, and we discuss
this in detail in Appendix \ref{sec:compare}. The distribution of the
best fitting inclinations is essentially consistent with the
inclinations being random ($i = 60\degr$), as the mean inclination
value is $\bar{i} = 55\pm25\degr$.  The temperature gradients are
skewed toward higher values, but have a mean of $\bar{p} = -0.6 \pm
0.5$, close to the $-0.75$ suggested for a flat accretion disc.  We
note that two objects show well constrained temperature gradients of
$-0.43$, which are consistent with flared, irradiated discs
\citep{chiang_1997}. The surface density gradients are more evenly
spread across the parameter space, with a mean of $\bar{q} = -1.5 \pm
1.2$.  This is consistent with the surface density gradient for a flat
accretion disc, although it is associated with a large error. The
average intrinsic linewidth of the fits is $\bar{\Delta \nu} = 20\pm
12$\,\kms. Apart from two objects (G287.3716$+$00.6444 and IRAS
17441$-$2910) the linewidths are approximately ten times the thermal
linewidths expected for CO between 1000-5000K, suggesting that the
emitting material is dominated by macro-turbulent motions or infall.

\section{Discussion}
\label{sec:discussion}

\subsection{Disc sizes - the location of the emission}

We find a range of disc sizes from our results. However, in the
majority of cases, the inner edge of the CO emission region is within
a few au.  The outer edge of the emission region varies much more,
giving rise to three very large discs.  However, the objects with very
large discs have temperature exponents that are poorly constrained.
This allows very shallow gradients (as seen in Figure
\ref{fig:incpq_histo}), which in turn produce large outer disc radii,
because this is defined as where the excitation temperature drops
below 1000\,K.  Thus, these large disc sizes are likely not physical
and much smaller discs can be produced with an exponent that is still
within the uncertainty in the best fitting value.

\smallskip

\begin{figure}
\centering
\includegraphics[width=\columnwidth, angle=0,trim=1cm 1cm 1cm 2cm, clip=true]{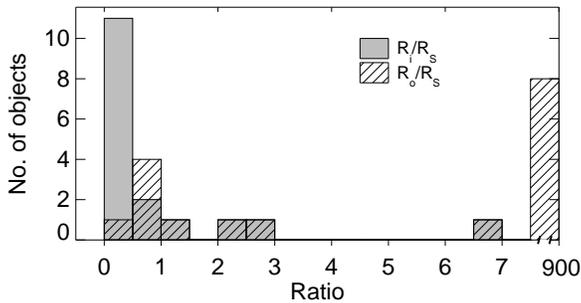}
\caption{Ratio of inner and outer CO disc radii, $R_{\mathrm{i}}$,
  $R_{\mathrm{o}}$, compared to the dust sublimation radius
  $R_{\mathrm{S}}$ for each object.  The majority of objects have
  inner radius well below the dust sublimation radius, and in some
  cases the entire CO disc is inside the sublimation radius. Most of
  the discs are within a few dust sublimation radii. Note that the
  final bin in the histogram is uneven and extends from 7.5--900.}
\label{fig:ri_rsub}
\end{figure}

To compare the location of the CO emission to the circumstellar disc
as a whole, we can calculate the dust sublimation radius,
$R_{\mathrm{S}}$, in au for each object:
\begin{equation}
R_{\mathrm{S}} = 1.1 \sqrt{Q_{\mathrm{R}}} \left( \frac{ L_{\star}}{1000\,L_{\odot}}\right)^{0.5} \left(\frac{T_{\mathrm{S}}}{1500\,\mathrm{K}}\right)^{-2},
\end{equation}
where $Q_{\mathrm{R}}$ is the ratio of absorption efficiencies of the
dust, and $T_{\mathrm{S}}$ is the temperature at which the dust
sublimates \citep{monnier_2002}.  We take $Q_{\mathrm{R}} = 1$ and
$T_{\mathrm{S}} = 1500$\,K.  This is then compared to both the inner
and outer radii of the CO emission disc from our model fits.  Figure
\ref{fig:ri_rsub} shows histograms of the ratios between these
quantities.

\smallskip

As expected, we find that the majority of objects, approximately 75
per cent, have CO discs whose inner extent is less than the dust
sublimation radius. Approximately 30 per cent have CO discs with an
outer extent below the dust sublimation radius. However this
percentage is increased if we exclude the artificially large discs
discussed earlier. Concerning the remaining object, there are several
cases where the inner and/or outer disc radii are only a few times
larger than the dust sublimation radius. Our treatment of the
sublimation is simplistic. Factors such as rapid accretion rates,
back-warming and non-homogeneous dust grain sizes may increase the dust
sublimation radius to several times the value we predict.  Therefore,
the best fitting disc properties appear consistent with physical
expectations. Consequently, we suggest that, in general, the best
fitting discs can be associated with gaseous accretion discs that are
close to the central star. As this is the largest sample of MYSOs with
CO bandhead emission studied to date, this provides strong evidence
for the existence of small scale accretion discs around these objects.

\subsection{Determining mass accretion rates}

As we discuss previously, the use of a physical disc model that
directly includes the relevant physics needed to determine a mass
accretion rate is attractive. However, as these models did not fit our
data well, we were unable to find accretion rates directly from the
fits.  Therefore, we investigated comparing the radial profiles of the
temperature and density from our fits to those predicted by an
accretion disc (model C) for $\dot{M} =
10^{-3.5}$--$10^{-7.5}$\,\msunyr, in an effort to determine an
estimate for the mass accretion rate of each object. We find that, for
many of the objects, the mass accretion rates obtained from the
temperature profile and the density profile differed by orders of
magnitude. Also, several accretion rates were difficult to determine
due to very different gradients between our fit and the accretion disc
model, and the values obtained are therefore unreliable.  On average,
we found that the mass accretion rate from the temperature gradients
suggested $\dot{M} > 10^{-4}\,$\msunyr\, while the surface density
distribution suggested $\dot{M} < 10^{-7}\,$\msunyr.

\smallskip

We conclude that our model fits cannot currently be used to determine
the mass accretion rates of MYSOs.  While this has been performed in
previous studies of lower mass young stellar objects \citep{carr_1989,
  chandler_1995}, the higher resolution of our data demonstrates there
are many rotational lines that cannot be fitted using these simple disc
models. These models may be inadequate in accurately representing the
physical situation that gives rise to CO bandhead emission from discs
with high accretion rates. Also, the study of only a single bandhead
may not offer sufficient information to reliably determine the
physical properties of discs and thus cannot accurately constrain mass
accretion rates in this way.  Observations covering a larger range in
wavelength, for instance with VLT/XSHOOTER, would allow fitting of
additional bandheads that probe different temperature regimes, at the
expense of the high spectral resolution employed in this paper.

\subsection{Are these typical MYSOs?}

In this section we consider whether the MYSOs with CO emission differ
from those without. An important question since the detection rate of
CO emission in the spectra of MYSOs is approximately 25 per cent.

\smallskip

MYSOs are typically red objects with a featureless continuum in the
optical and NIR, so we cannot accurately constrain their stellar
properties such as effective temperature and radius. Therefore, we
investigate whether the parameters of the best fitting models display
a pattern which may explain why only some MYSOs exhibit this
emission. In particular, it is conceivable that for CO emission to be
observed, the inclination of the disc is required to be close to face
on.  However, the inclinations of the objects are spread relatively
evenly between 0 and $90\degr$, with a slight preference for higher
inclinations, suggesting this is not the case. In addition, we note
that the average bolometric luminosity of our sample of objects with
CO emission is $5\times 10^{4}\,\mathrm{L}_{\odot}$, which is typical
of objects within the RMS survey. Therefore, we conclude that the
properties of the MYSOs that exhibit CO bandhead emission do not
indicate what specific geometry and/or conditions result are required
for the presence of this emission.

\smallskip

\begin{figure}
\centering
\includegraphics[width=\columnwidth]{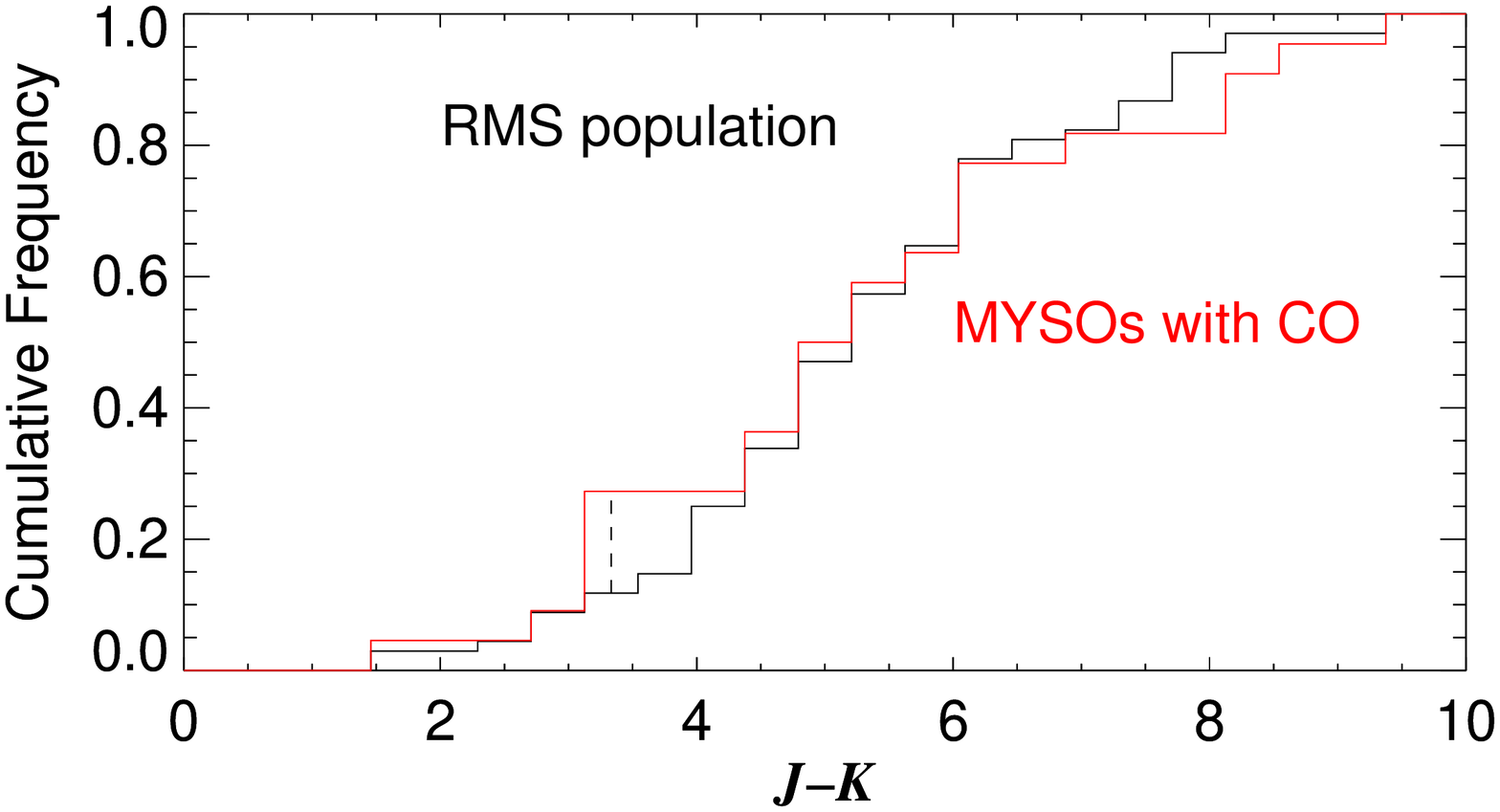}
\includegraphics[width=\columnwidth]{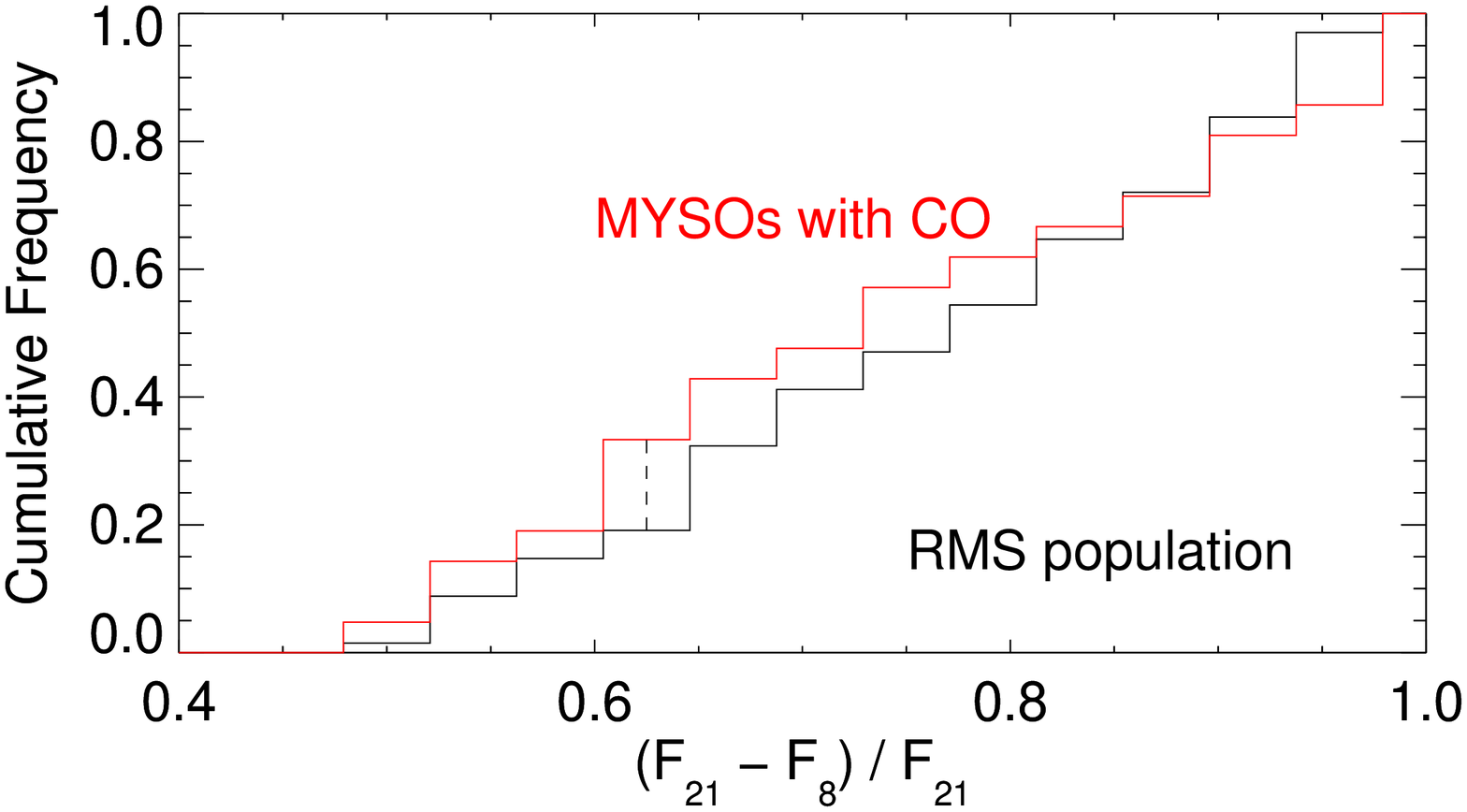}
\caption{Cumulative frequency of the NIR $J-K$ and normalised MIR
$(F_{21\mathrm{\mu m}}-F_{8\mathrm{\mu m}})$ colours of the MYSOs with
CO emission and the RMS population.  The dashed vertical lines
indicate the largest deviation between the distributions.
Similar distributions were found for the $H-K$, $(F_{21\mathrm{\mu
m}}-F_{12\mathrm{\mu m}})$ and $(F_{21\mathrm{\mu m}}-F_{14\mathrm{\mu
m}})$ colours.}
\label{fig:mir}
\end{figure}

Consequently, it is not certain why the presence of CO bandhead
emission is not ubiquitous in the spectra of MYSOs. It is possible
that the objects with CO emission represent a different evolutionary
stage of MYSOs than those without. It is difficult to test this
hypothesis. As an initial test, we investigate the infrared colours of
the observed MYSOs, which are likely affected by key factors such as
circumstellar geometries on astronomical unit scales, inclination and
envelope mass/infall rate. To determine whether the MYSOs with CO
bandhead emission appear representative of MYSOs in general or are a
specific subset of MYSOs, we compared their NIR colours with those of
approximately 70 objects from the RMS database, which we show in the
upper panel of Figure \ref{fig:mir}.  To ensure a valid comparison
with our objects, the control sample was selected to have a high
luminosity ($L > 10^{4}\,L_{\odot}$) and be bright in the $K-$band ($K
< 10$ magnitudes).  We compared the $J-K$ and $H-K$ colours of the two
samples with the Kolmogorov--Smirnov (KS) test and found that the
hypothesis that the NIR colours of the MYSOs with CO bandhead emission
are drawn from the distribution of NIR colours exhibited by the RMS
population cannot be discounted with any significance. The objects
observed to have CO bandhead emission are generally bright in the
$K-$band. However, this is a selection effect and does not imply that
only objects with such emission are intrinsically brighter in the
$K-$band.

\smallskip

We also examined the mid--infrared (MIR) colours of the sample.
Specifically, we compared the \textit{MSX} $(F_{21\mathrm{\mu
m}}-F_{8\mathrm{\mu m}})$, $(F_{21\mathrm{\mu m}}-F_{12\mathrm{\mu
m}})$ and $(F_{21\mathrm{\mu m}}-F_{14\mathrm{\mu m}})$ colours of the
objects with CO bandhead emission to the RMS population mentioned
earlier.  The lower panel of Figure \ref{fig:mir} shows the cumulative
distribution of the $(F_{21\mathrm{\mu m}}-F_{8\mathrm{\mu m}})$
colours. Using the KS test, we found that the hypothesis that the MIR
colours of the objects with CO emission are drawn from the total RMS
distribution of MIR colours cannot be discounted with any
significance.  Therefore, the objects with and without CO emission
appear no different in terms of both their NIR and MIR colours,
suggesting these objects are representative of the RMS population as a
whole.  

\smallskip

It is therefore unclear from this analysis why only some MYSOs possess
CO emission.  It has been predicted that models of circumstellar discs
of MYSOs can be unstable \citep{krumholz_2007b} and that the accretion
rate in these discs is not constant \citep{kuiper_2011}.  Unstable
discs may disrupt the circumstellar material and lead to physical
conditions in which CO ro-vibrational emission no longer occurs.  If
the disc accretion rate is high, this will increase the mid-plane
temperature of the disc and move the CO emission region further away
from the central protostar, possibly into the surrounding envelope,
which may cause the emission to cease.  However, without a method to
determine the accretion rate of these objects, this hypothesis is
difficult to test.

\subsection{Discs and massive star formation}

The physical processes occurring during massive star formation are
still not well understood. It is thought accretion proceeds through
circumstellar discs at high rates, which are believed to affect the
state of the central protostar
\citep{hosokawa_2009,hosokawa_2010}. Furthermore, recent simulations
indicate that high accretion rates result in massive discs and that
gravitational torques in such self-gravitating discs provide a
mechanism to transport angular momentum outwards
\citep{kuiper_2011}. Therefore, it is becoming apparent that accretion
discs play a central role in the formation of massive stars. However,
it is difficult to confirm this observationally. In most cases, the
large distances to MYSOs, and their embedded nature, prevent
detections of astronomical unit scale circumstellar discs \citep[the
exception being IRAS 13481, see][]{kraus_2010}. In some cases, high
infall rates have been detected towards massive star forming regions
\citep[see
e.g.][]{beuther_2012,wyrowski_2012,qiu_2012,herpin_2012}. However,
these observations probe scales of approximately 1000\,au and
larger. Therefore, it is difficult to establish that this material
will be accreted by a single object. Consequently, our observational
overview of massive star formation is still incomplete.

\smallskip

With this in mind, we note that our sample constitutes the largest
sample of MYSOs with CO bandhead emission studied to date
\citep[several times that of][]{wheelwright_2010}. Therefore, by
confronting the observations with a kinematic model, we can conduct an
extensive investigation into the circumstellar environment of these
objects. We demonstrate that all the observed bandhead profiles can be
successfully fitted with a model of a circumstellar disc in Keplerian
rotation. In addition, we have shown that essentially all these models
can be associated with gaseous discs interior to the dust sublimation
radii of these objects. Finally, we demonstrate that the objects in
question appear no different to the ensemble of objects in the RMS
catalogue, the largest most complete catalogue of MYSOs to
date. Therefore, the fact that the CO bandhead emission of all the
objects observed can be fitted with a disc model supports the scenario in
which all MYSOs are surrounded by accretion discs. With our limited
wavelength coverage, we were not able to establish precise constraints
on the accretion rates. Nonetheless, we find that the disc properties
are consistent with high accretion rates. Therefore, these
observations are entirely consistent with the hypothesis that MYSOs
are massive stars in the process of forming that accrete matter
through circumstellar discs.

\section{Conclusions}
\label{sec:conclusions}

In this paper we have presented the near-infrared spectra of 20
massive young stellar objects that possess CO first overtone bandhead
emission, the largest sample studied in this way to date.  We have fit
the spectra of these objects with a model of emission originating in a
circumstellar disc in Keplerian rotation.  We tested three approaches
to describing the properties of such circumstellar discs and found
that the spectra were best fit by using an analytic approach to
describe the temperature and density within the disc.  Our main
findings are:

\begin{enumerate}

\item All spectra are well fit by a model of a Keplerian rotation
disc.

\item The best fitting disc parameters are consistent with
previously published information. The inclinations are spread across
a wide rage of angles. The best fitting temperature and density
exponents are, on average, consistent with flat circumstellar discs
(subject to some scatter), a handful of objects have exponents
consistent with flared, irradiated discs.

\item Essentially all the best fitting discs are located close to the
  dust sublimation radius, which is consistent with the existence of
  small scale gaseous accretion discs around these objects.

\end{enumerate}

We found that the mass accretion rates of the objects are not easily
determined from examination of the disc structure as traced by a
single CO bandhead.  The analysis of further bandheads may allow these
accretion rates to be determined.  Our high spectral resolution
observations of the CO emission of a substantial number of MYSOs and
our modelling of this emission are entirely consistent with the
scenario in which MYSOs are surrounded by small scale ($<100\,$au) accretion
discs. The objects observed constitute a large sample of circumstellar
discs around MYSOs and provide promising targets for future
observations and inspiration for detailed modelling of the accretion
environment of such objects.

\section*{Acknowledgments}

We would like to thank Bhargav Vaidya for useful discussions regarding
the physical disc models, and the anonymous referee whose
thorough comments helped to improve the paper.  JDI and LTM
gratefully acknowledge studentships from the Science and Technology
Facilities Council of the United Kingdom (STFC). HEW acknowledges the
financial support of the MPIfR in Bonn. This paper has made use of
information from the RMS survey database at
http://www.ast.leeds.ac.uk/RMS which was constructed with support from
the Science and Technology Facilities Council of the United Kingdom.

\bibliographystyle{mn2e_long}

\begin{thebibliography}{71}

\bibitem[{{Berthoud} {et~al}\mbox{.}(2007){Berthoud}, {Keller}, {Herter},
  {Richter}, \& {Whelan}}]{berthoud_2007}
{Berthoud} M.~G., {Keller} L.~D., {Herter} T.~L., {Richter} M.~J., {Whelan}
  D.~G., 2007, \apj, 660, 461

\bibitem[{{Beuther} {et~al}\mbox{.}(2012){Beuther}, {Linz}, \&
  {Henning}}]{beuther_2012}
{Beuther} H., {Linz} H., {Henning} T., 2012, \aap, 543, A88

\bibitem[{{Bik} {et~al}\mbox{.}(2006){Bik}, {Kaper}, \& {Waters}}]{bik_2006}
{Bik} A., {Kaper} L., {Waters} L.~B.~F.~M., 2006, \aap, 455, 561

\bibitem[{{Bik} \& {Thi}(2004)}]{bik_2004}
{Bik} A., {Thi} W.~F., 2004, \aap, 427, L13

\bibitem[{{Blum} {et~al}\mbox{.}(2004){Blum}, {Barbosa}, {Damineli}, {Conti},
  \& {Ridgway}}]{blum_2004}
{Blum} R.~D., {Barbosa} C.~L., {Damineli} A., {Conti} P.~S., {Ridgway} S.,
  2004, \apj, 617, 1167

\bibitem[{{Boley} {et~al}\mbox{.}(2012){Boley}, {Linz}, {van Boekel},
  {Bouwman}, {Henning}, \& {Sobolev}}]{pboley_2012}
{Boley} P., {Linz} H., {van Boekel} R., {Bouwman} J., {Henning} T., {Sobolev}
  A., 2012, ArXiv e-prints

\bibitem[{{Brand} \& {Blitz}(1993)}]{brand_1993}
{Brand} J., {Blitz} L., 1993, \aap, 275, 67

\bibitem[{{Bronfman} {et~al}\mbox{.}(1996){Bronfman}, {Nyman}, \&
  {May}}]{bronfman_1996}
{Bronfman} L., {Nyman} L.-A., {May} J., 1996, \aaps, 115, 81

\bibitem[{{Carr}(1989)}]{carr_1989}
{Carr} J.~S., 1989, \apj, 345, 522

\bibitem[{{Carrasco-Gonz{\'a}lez} {et~al}\mbox{.}(2012){Carrasco-Gonz{\'a}lez},
  {Galv{\'a}n-Madrid}, {Anglada}, {Osorio}, {D'Alessio}, {Hofner},
  {Rodr{\'{\i}}guez}, {Linz}, \& {Araya}}]{carrasco_2012}
{Carrasco-Gonz{\'a}lez} C. {et~al.}, 2012, \apjl, 752, L29

\bibitem[{{Chandler} {et~al}\mbox{.}(1995){Chandler}, {Carlstrom}, \&
  {Scoville}}]{chandler_1995}
{Chandler} C.~J., {Carlstrom} J.~E., {Scoville} N.~Z., 1995, \apj, 446, 793

\bibitem[{{Chiang} \& {Goldreich}(1997)}]{chiang_1997}
{Chiang} E.~I., {Goldreich} P., 1997, \apj, 490, 368

\bibitem[{{Chini} \& {Neckel}(1981)}]{chini_1981}
{Chini} R., {Neckel} T., 1981, \aap, 102, 171

\bibitem[{{{Clarke}, A.~J.}(2007)}]{clarke_phd_2007}
{{Clarke}, A.~J.}, 2007, PhD thesis, {University of Leeds}

\bibitem[{Cooper(in press.)}]{cooper_prep}
Cooper H., {in press}., \mnras

\bibitem[{{Covey} {et~al}\mbox{.}(2011){Covey}, {Hillenbrand}, {Miller},
  {Poznanski}, {Cenko}, {Silverman}, {Bloom}, {Kasliwal}, {Fischer}, {Rayner},
  {Rebull}, {Butler}, {Filippenko}, {Law}, {Ofek}, {Ag{\"u}eros}, {Dekany},
  {Rahmer}, {Hale}, {Smith}, {Quimby}, {Nugent}, {Jacobsen}, {Zolkower},
  {Velur}, {Walters}, {Henning}, {Bui}, {McKenna}, {Kulkarni}, \&
  {Klein}}]{covey_2011}
{Covey} K.~R. {et~al.}, 2011, \aj, 141, 40

\bibitem[{{Davies} {et~al}\mbox{.}(2010){Davies}, {Lumsden}, {Hoare},
  {Oudmaijer}, \& {de Wit}}]{davies_2010}
{Davies} B., {Lumsden} S.~L., {Hoare} M.~G., {Oudmaijer} R.~D., {de Wit} W.-J.,
  2010, \mnras, 402, 1504

\bibitem[{{de Wit} {et~al}\mbox{.}(2010){de Wit}, {Hoare}, {Oudmaijer}, \&
  {Lumsden}}]{deWit_2010}
{de Wit} W.~J., {Hoare} M.~G., {Oudmaijer} R.~D., {Lumsden} S.~L., 2010, \aap,
  515, A45

\bibitem[{{Egan} {et~al}\mbox{.}(2003){Egan}, {Price}, \&
  {Kraemer}}]{egan_2003}
{Egan} M.~P., {Price} S.~D., {Kraemer} K.~E., 2003, in Bulletin of the AAS,
  Vol.~35, AAS Meeting Abstracts, p. 1301

\bibitem[{{Ellerbroek} {et~al}\mbox{.}(2011){Ellerbroek}, {Kaper}, {Bik}, {de
  Koter}, {Horrobin}, {Puga}, {Sana}, \& {Waters}}]{ellerbroek_2011}
{Ellerbroek} L.~E., {Kaper} L., {Bik} A., {de Koter} A., {Horrobin} M., {Puga}
  E., {Sana} H., {Waters} L.~B.~F.~M., 2011, \apjl, 732, L9

\bibitem[{{Ferguson} {et~al}\mbox{.}(2005){Ferguson}, {Alexander}, {Allard},
  {Barman}, {Bodnarik}, {Hauschildt}, {Heffner-Wong}, \&
  {Tamanai}}]{ferguson_2005}
{Ferguson} J.~W., {Alexander} D.~R., {Allard} F., {Barman} T., {Bodnarik}
  J.~G., {Hauschildt} P.~H., {Heffner-Wong} A., {Tamanai} A., 2005, \apj, 623,
  585

\bibitem[{{Herpin} {et~al}\mbox{.}(2012){Herpin}, {Chavarr{\'{\i}}a}, {van der
  Tak}, {Wyrowski}, {van Dishoeck}, {Jacq}, {Braine}, {Baudry}, {Bontemps}, \&
  {Kristensen}}]{herpin_2012}
{Herpin} F. {et~al.}, 2012, \aap, 542, A76

\bibitem[{{Hosokawa} \& {Omukai}(2009)}]{hosokawa_2009}
{Hosokawa} T., {Omukai} K., 2009, \apj, 691, 823

\bibitem[{{Hosokawa} {et~al}\mbox{.}(2010){Hosokawa}, {Yorke}, \&
  {Omukai}}]{hosokawa_2010}
{Hosokawa} T., {Yorke} H.~W., {Omukai} K., 2010, \apj, 721, 478

\bibitem[{{Jim{\'e}nez-Serra} {et~al}\mbox{.}(2007){Jim{\'e}nez-Serra},
  {Mart{\'{\i}}n-Pintado}, {Rodr{\'{\i}}guez-Franco}, {Chandler}, {Comito}, \&
  {Schilke}}]{jiminez_2007}
{Jim{\'e}nez-Serra} I., {Mart{\'{\i}}n-Pintado} J., {Rodr{\'{\i}}guez-Franco}
  A., {Chandler} C., {Comito} C., {Schilke} P., 2007, \apjl, 661, L187

\bibitem[{{Kaeufl} {et~al}\mbox{.}(2004){Kaeufl}, {Ballester}, {Biereichel},
  {Delabre}, {Donaldson}, {Dorn}, {Fedrigo}, \& {Finger}}]{kaeuful_2004}
{Kaeufl} H.-U., {Ballester} P., {Biereichel} P., {Delabre} B., {Donaldson} R.,
  {Dorn} R., {Fedrigo} E., {Finger}, 2004, in SPIE Conference Series, Vol.
  5492, SPIE Conference Series, {Moorwood} A.~F.~M., {Iye} M., eds., pp.
  1218--1227

\bibitem[{{Kahn}(1974)}]{kahn_1974}
{Kahn} F.~D., 1974, \aap, 37, 149

\bibitem[{{Klaassen} {et~al}\mbox{.}(2011){Klaassen}, {Wilson}, {Keto},
  {Zhang}, {Galv{\'a}n-Madrid}, \& {Liu}}]{klaassen_2011}
{Klaassen} P.~D., {Wilson} C.~D., {Keto} E.~R., {Zhang} Q., {Galv{\'a}n-Madrid}
  R., {Liu} H.-Y.~B., 2011, \aap, 530, A53

\bibitem[{{Kraus} {et~al}\mbox{.}(2000){Kraus}, {Kr{\"u}gel}, {Thum}, \&
  {Geballe}}]{kraus_2000}
{Kraus} M., {Kr{\"u}gel} E., {Thum} C., {Geballe} T.~R., 2000, \aap, 362, 158

\bibitem[{{Kraus} {et~al}\mbox{.}(2010){Kraus}, {Hofmann}, {Menten}, {Schertl},
  {Weigelt}, {Wyrowski}, {Meilland}, {Perraut}, {Petrov}, {Robbe-Dubois},
  {Schilke}, \& {Testi}}]{kraus_2010}
{Kraus} S. {et~al.}, 2010, \nat, 466, 339

\bibitem[{{Krumholz} {et~al}\mbox{.}(2007){Krumholz}, {Klein}, \&
  {McKee}}]{krumholz_2007b}
{Krumholz} M.~R., {Klein} R.~I., {McKee} C.~F., 2007, \apj, 665, 478

\bibitem[{{Krumholz} {et~al}\mbox{.}(2009){Krumholz}, {Klein}, {McKee},
  {Offner}, \& {Cunningham}}]{krumholz_2009}
{Krumholz} M.~R., {Klein} R.~I., {McKee} C.~F., {Offner} S.~S.~R., {Cunningham}
  A.~J., 2009, Science, 323, 754

\bibitem[{{Kuiper} {et~al}\mbox{.}(2010){Kuiper}, {Klahr}, {Beuther}, \&
  {Henning}}]{kuiper_2010}
{Kuiper} R., {Klahr} H., {Beuther} H., {Henning} T., 2010, \apj, 722, 1556

\bibitem[{{Kuiper} {et~al}\mbox{.}(2011){Kuiper}, {Klahr}, {Beuther}, \&
  {Henning}}]{kuiper_2011}
{Kuiper} R., {Klahr} H., {Beuther} H., {Henning} T., 2011, \apj, 732, 20

\bibitem[{{Larson} \& {Starrfield}(1971)}]{larson_1971}
{Larson} R.~B., {Starrfield} S., 1971, \aap, 13, 190

\bibitem[{{Linz} {et~al}\mbox{.}(2009){Linz}, {Henning}, {Feldt}, {Pascucci},
  {van Boekel}, {Men'shchikov}, {Stecklum}, {Chesneau}, {Ratzka}, {Quanz},
  {Leinert}, {Waters}, \& {Zinnecker}}]{linz_2009}
{Linz} H. {et~al.}, 2009, \aap, 505, 655

\bibitem[{{Lumsden} {et~al}\mbox{.}(2002){Lumsden}, {Hoare}, {Oudmaijer}, \&
  {Richards}}]{lumsden_2002}
{Lumsden} S.~L., {Hoare} M.~G., {Oudmaijer} R.~D., {Richards} D., 2002, \mnras,
  336, 621

\bibitem[{{Martin}(1997)}]{martin_1997}
{Martin} S.~C., 1997, \apjl, 478, L33

\bibitem[{{Martins} {et~al}\mbox{.}(2005){Martins}, {Schaerer}, \&
  {Hillier}}]{martins_2005}
{Martins} F., {Schaerer} D., {Hillier} D.~J., 2005, \aap, 436, 1049

\bibitem[{{Molinari} {et~al}\mbox{.}(1996){Molinari}, {Brand}, {Cesaroni}, \&
  {Palla}}]{molinari_1996}
{Molinari} S., {Brand} J., {Cesaroni} R., {Palla} F., 1996, \aap, 308, 573

\bibitem[{{Monnier} \& {Millan-Gabet}(2002)}]{monnier_2002}
{Monnier} J.~D., {Millan-Gabet} R., 2002, \apj, 579, 694

\bibitem[{{Mottram} {et~al}\mbox{.}(2011{\natexlab{a}}){Mottram}, {Hoare},
  {Davies}, {Lumsden}, {Oudmaijer}, {Urquhart}, {Moore}, {Cooper}, \&
  {Stead}}]{mottram_2011b}
{Mottram} J.~C. {et~al.}, 2011{\natexlab{a}}, \apjl, 730, L33

\bibitem[{{Mottram} {et~al}\mbox{.}(2010){Mottram}, {Hoare}, {Lumsden},
  {Oudmaijer}, {Urquhart}, {Meade}, {Moore}, \& {Stead}}]{mottram_2010}
{Mottram} J.~C., {Hoare} M.~G., {Lumsden} S.~L., {Oudmaijer} R.~D., {Urquhart}
  J.~S., {Meade} M.~R., {Moore} T.~J.~T., {Stead} J.~J., 2010, \aap, 510, A89

\bibitem[{{Mottram} {et~al}\mbox{.}(2007){Mottram}, {Hoare}, {Lumsden},
  {Oudmaijer}, {Urquhart}, {Sheret}, {Clarke}, \& {Allsopp}}]{mottram_2007}
{Mottram} J.~C., {Hoare} M.~G., {Lumsden} S.~L., {Oudmaijer} R.~D., {Urquhart}
  J.~S., {Sheret} T.~L., {Clarke} A.~J., {Allsopp} J., 2007, \aap, 476, 1019

\bibitem[{{Mottram} {et~al}\mbox{.}(2011{\natexlab{b}}){Mottram}, {Hoare},
  {Urquhart}, {Lumsden}, {Oudmaijer}, {Robitaille}, {Moore}, {Davies}, \&
  {Stead}}]{mottram_2011a}
{Mottram} J.~C. {et~al.}, 2011{\natexlab{b}}, \aap, 525, A149

\bibitem[{{Norberg} \& {Maeder}(2000)}]{norberg_2000}
{Norberg} P., {Maeder} A., 2000, \aap, 359, 1025

\bibitem[{{Patel} {et~al}\mbox{.}(2005){Patel}, {Curiel}, {Sridharan}, {Zhang},
  {Hunter}, {Ho}, {Torrelles}, {Moran}, {G{\'o}mez}, \& {Anglada}}]{patel_2005}
{Patel} N.~A. {et~al.}, 2005, \nat, 437, 109

\bibitem[{{Petrov} {et~al}\mbox{.}(2007){Petrov}, {Malbet}, {Weigelt},
  {Antonelli}, {Beckmann}, {Bresson}, {Chelli}, {Dugu{\'e}}, {Duvert},
  {Gennari}, {Gl{\"u}ck}, {Kern}, {Lagarde}, {Le Coarer}, {Lisi}, {Millour},
  {Perraut}, {Puget}, {Rantakyr{\"o}}, {Robbe-Dubois}, {Roussel}, {Salinari},
  {Tatulli}, {Zins}, {Accardo}, {Acke}, {Agabi}, {Altariba}, {Arezki},
  {Aristidi}, {Baffa}, {Behrend}, {Bl{\"o}cker}, {Bonhomme}, {Busoni},
  {Cassaing}, {Clausse}, {Colin}, {Connot}, {Delboulb{\'e}}, {Domiciano de
  Souza}, {Driebe}, {Feautrier}, {Ferruzzi}, {Forveille}, {Fossat}, {Foy},
  {Fraix-Burnet}, {Gallardo}, {Giani}, {Gil}, {Glentzlin}, {Heiden},
  {Heininger}, {Hernandez Utrera}, {Hofmann}, {Kamm}, {Kiekebusch}, {Kraus},
  {Le Contel}, {Le Contel}, {Lesourd}, {Lopez}, {Lopez}, {Magnard}, {Marconi},
  {Mars}, {Martinot-Lagarde}, {Mathias}, {M{\`e}ge}, {Monin}, {Mouillet},
  {Mourard}, {Nussbaum}, {Ohnaka}, {Pacheco}, {Perrier}, {Rabbia}, {Rebattu},
  {Reynaud}, {Richichi}, {Robini}, {Sacchettini}, {Schertl}, {Sch{\"o}ller},
  {Solscheid}, {Spang}, {Stee}, {Stefanini}, {Tallon}, {Tallon-Bosc}, {Tasso},
  {Testi}, {Vakili}, {von der L{\"u}he}, {Valtier}, {Vannier}, \&
  {Ventura}}]{petrov_2007}
{Petrov} R.~G. {et~al.}, 2007, \aap, 464, 1

\bibitem[{{Qiu} {et~al}\mbox{.}(2012){Qiu}, {Zhang}, {Beuther}, \&
  {Fallscheer}}]{qiu_2012}
{Qiu} K., {Zhang} Q., {Beuther} H., {Fallscheer} C., 2012, \apj, 756, 170

\bibitem[{{Robitaille} {et~al}\mbox{.}(2007){Robitaille}, {Whitney},
  {Indebetouw}, \& {Wood}}]{robitaille_2007}
{Robitaille} T.~P., {Whitney} B.~A., {Indebetouw} R., {Wood} K., 2007, \apjs,
  169, 328

\bibitem[{{Rothman} {et~al}\mbox{.}(1998){Rothman}, {Rinsland}, {Goldman},
  {Massie}, {Edwards}, {Flaud}, \& {Perrin}}]{rothman_1998}
{Rothman} L.~S., {Rinsland} C.~P., {Goldman} A., {Massie} S.~T., {Edwards}
  D.~P., {Flaud} J.-M., {Perrin} A., 1998, \jqsrt, 60, 665

\bibitem[{{Shakura} \& {Sunyaev}(1973)}]{shakura_1973}
{Shakura} N.~I., {Sunyaev} R.~A., 1973, \aap, 24, 337

\bibitem[{{Shu} {et~al}\mbox{.}(1987){Shu}, {Adams}, \& {Lizano}}]{shu_1987}
{Shu} F.~H., {Adams} F.~C., {Lizano} S., 1987, \araa, 25, 23

\bibitem[{{Sridharan} {et~al}\mbox{.}(2002){Sridharan}, {Beuther}, {Schilke},
  {Menten}, \& {Wyrowski}}]{sridharan_2002}
{Sridharan} T.~K., {Beuther} H., {Schilke} P., {Menten} K.~M., {Wyrowski} F.,
  2002, \apj, 566, 931

\bibitem[{{Tatulli} {et~al}\mbox{.}(2008){Tatulli}, {Malbet}, {M{\'e}nard},
  {Gil}, {Testi}, {Natta}, {Kraus}, {Stee}, \& {Robbe-Dubois}}]{tatulli_2008}
{Tatulli} E. {et~al.}, 2008, \aap, 489, 1151

\bibitem[{{Urquhart} {et~al}\mbox{.}(2007{\natexlab{a}}){Urquhart}, {Busfield},
  {Hoare}, {Lumsden}, {Clarke}, {Moore}, {Mottram}, \&
  {Oudmaijer}}]{urquhart_2007a}
{Urquhart} J.~S., {Busfield} A.~L., {Hoare} M.~G., {Lumsden} S.~L., {Clarke}
  A.~J., {Moore} T.~J.~T., {Mottram} J.~C., {Oudmaijer} R.~D.,
  2007{\natexlab{a}}, \aap, 461, 11

\bibitem[{{Urquhart} {et~al}\mbox{.}(2007{\natexlab{b}}){Urquhart}, {Busfield},
  {Hoare}, {Lumsden}, {Oudmaijer}, {Moore}, {Gibb}, {Purcell}, {Burton}, \&
  {Marechal}}]{urquhart_2007b}
{Urquhart} J.~S. {et~al.}, 2007{\natexlab{b}}, \aap, 474, 891

\bibitem[{{Urquhart} {et~al}\mbox{.}(2008){Urquhart}, {Busfield}, {Hoare},
  {Lumsden}, {Oudmaijer}, {Moore}, {Gibb}, {Purcell}, {Burton}, {Mar{\'e}chal},
  {Jiang}, \& {Wang}}]{urquhart_2008}
{Urquhart} J.~S. {et~al.}, 2008, \aap, 487, 253

\bibitem[{{Urquhart} {et~al}\mbox{.}(2009{\natexlab{a}}){Urquhart}, {Hoare},
  {Lumsden}, {Oudmaijer}, {Moore}, {Brook}, {Mottram}, {Davies}, \&
  {Stead}}]{urquhart_2009a}
{Urquhart} J.~S. {et~al.}, 2009{\natexlab{a}}, \aap, 507, 795

\bibitem[{{Urquhart} {et~al}\mbox{.}(2012){Urquhart}, {Hoare}, {Lumsden},
  {Oudmaijer}, {Moore}, {Mottram}, {Cooper}, {Mottram}, \&
  {Rogers}}]{urquhart_2012}
{Urquhart} J.~S. {et~al.}, 2012, \mnras, 420, 1656

\bibitem[{{Urquhart} {et~al}\mbox{.}(2011){Urquhart}, {Moore}, {Hoare},
  {Lumsden}, {Oudmaijer}, {Rathborne}, {Mottram}, {Davies}, \&
  {Stead}}]{urquhart_2011}
{Urquhart} J.~S. {et~al.}, 2011, \mnras, 410, 1237

\bibitem[{{Urquhart} {et~al}\mbox{.}(2009{\natexlab{b}}){Urquhart}, {Morgan},
  \& {Thompson}}]{urquhart_2009b}
{Urquhart} J.~S., {Morgan} L.~K., {Thompson} M.~A., 2009{\natexlab{b}}, \aap,
  497, 789

\bibitem[{{Vaidya} {et~al}\mbox{.}(2009){Vaidya}, {Fendt}, \&
  {Beuther}}]{vaidya_2009}
{Vaidya} B., {Fendt} C., {Beuther} H., 2009, \apj, 702, 567

\bibitem[{{van Dishoeck} \& {Black}(1988)}]{vandishoeck_1988}
{van Dishoeck} E.~F., {Black} J.~H., 1988, \apj, 334, 771

\bibitem[{{Walsh} {et~al}\mbox{.}(1997){Walsh}, {Hyland}, {Robinson}, \&
  {Burton}}]{walsh_1997}
{Walsh} A.~J., {Hyland} A.~R., {Robinson} G., {Burton} M.~G., 1997, \mnras,
  291, 261

\bibitem[{{Wheelwright} {et~al}\mbox{.}(2012{\natexlab{a}}){Wheelwright}, {de
  Wit}, {Oudmaijer}, {Hoare}, {Lumsden}, {Fujiyoshi}, \&
  {Close}}]{wheelwright_2012_visir}
{Wheelwright} H.~E., {de Wit} W.~J., {Oudmaijer} R.~D., {Hoare} M.~G.,
  {Lumsden} S.~L., {Fujiyoshi} T., {Close} J.~L., 2012{\natexlab{a}}, \aap,
  540, A89

\bibitem[{{Wheelwright} {et~al}\mbox{.}(2012{\natexlab{b}}){Wheelwright}, {de
  Wit}, {Weigelt}, {Oudmaijer}, \& {Ilee}}]{wheelwright_2012_amber}
{Wheelwright} H.~E., {de Wit} W.~J., {Weigelt} G., {Oudmaijer} R.~D., {Ilee}
  J.~D., 2012{\natexlab{b}}, \aap, 543, A77

\bibitem[{{Wheelwright} {et~al}\mbox{.}(2010){Wheelwright}, {Oudmaijer}, {de
  Wit}, {Hoare}, {Lumsden}, \& {Urquhart}}]{wheelwright_2010}
{Wheelwright} H.~E., {Oudmaijer} R.~D., {de Wit} W.~J., {Hoare} M.~G.,
  {Lumsden} S.~L., {Urquhart} J.~S., 2010, \mnras, 408, 1840

\bibitem[{{Wolfire} \& {Cassinelli}(1987)}]{wolfire_1987}
{Wolfire} M.~G., {Cassinelli} J.~P., 1987, \apj, 319, 850

\bibitem[{{Wyrowski} {et~al}\mbox{.}(2012){Wyrowski}, {G{\"u}sten}, {Menten},
  {Wiesemeyer}, \& {Klein}}]{wyrowski_2012}
{Wyrowski} F., {G{\"u}sten} R., {Menten} K.~M., {Wiesemeyer} H., {Klein} B.,
  2012, \aap, 542, L15

\bibitem[{{Zinnecker} \& {Yorke}(2007)}]{zinnecker_2007}
{Zinnecker} H., {Yorke} H.~W., 2007, \araa, 45, 481

\end{thebibliography}

\appendix

\section{Notes on individual objects \& comparison of results with previous studies}
\label{sec:compare}

Here we discuss and compare, on an object by object basis, our
findings with that of previous studies on a selection of our sample
where data was available.

\subsection{G033.3891$+$00.1989}

The work of \citet{wheelwright_2010} fitted the CO bandhead emission of
this object using fixed power law relations for the disc.  They found
a disc from 0.24--2.0\,au, at an inclination of $18\degr$, with an
intrinsic linewidth of 20\,\kms (which was not well constrained) and a
CO number density of $9\times 10^{21}$\,\cm2.  The inclination, number
density and linewidth agree with our best fitting model within error
ranges, but the size of our disc is much larger, due to a shallow
temperature gradient.

\subsection{G287.3716$-$00.6444}

\citet{wheelwright_2010} do not find a satisfactory fit to the CO
bandhead of this object assuming a circumstellar disc (with fixed
temperature and density profiles), nor an isothermal non-rotating body
of CO.  They discuss other possible origins for the emission,
including a disc with an outer bulge, a dense neutral wind, a shock,
or a disc in which the receding side is much brighter than the
approaching side.  We note that our temperature exponent is close to
$-0.43$, which would be consistent with a flared, irradiated disc
\citep{chiang_1997} which may act in the same way as a disc with an
outer bulge.

\subsection{G308.9176$+$00.1231 (AFGL\,4176)}

\citet{wheelwright_2010} find a disc from 1--8\,au at an inclination
of $30\degr$.  The size of the disc agrees well with our results, but we
find a higher inclination of $67\degr$, with $30\degr$ at the lower limit
of our error range.  Our linewidth of 12.6\,\kms agrees with their
value of 14\,\kms, however our inner density is one order of magnitude
lower.  \citet{pboley_2012} find their observations described well by
a large circumstellar disc at an inclination of $60\degr$, agreeing with
our best fitting model, and consistent with the prominent blue
shoulder in our data.

\subsection{G310.0135$+$00.3892}

G310.0135$+$00.3892 (IRAS 13481$-$6124) has previously been observed
in the $K-$band by \citet{kraus_2010} who report an elongated
structure that is consistent with a disc viewed at a moderate
inclination of approximately $45\degr$.  \citet{wheelwright_2012_visir}
fitted the SED using a model with an inclination of $32\degr$.  Our
relatively high inclination of $67\degr$ is not well constrained, due to
the poor quality of the data, but agrees with these values within the
error range.

\smallskip

\citet{kraus_2010} find a temperature gradient of $p \sim -0.4$,
which they suggest is consistent with a flared, irradiated disc based
on the work of \citet{chiang_1997}.  We find a temperature gradient of
$p = -0.43$, which is consistent with this hypothesis.  Our inner disc
temperature of $3800\,$K is warmer than the value of approximately
1500--2000\,K assumed in \citet{kraus_2010}, but is consistent as we
are concerned with a gaseous disc as opposed to a dust disc. We find a
smaller inner radius ($2.8\,$au) for our disc than their study
($9.5\,$ au).

\subsection{G332.9868$-$00.4871}

\citet{wheelwright_2012_visir} determine an inclination of $15\degr$
to G332.9868$-$00.4871, which is far from our reported value of
$78\degr$ even considering the large error.  However, we note that our
data have a relatively low signal-to-noise ratio, and no rotational
lines in the fourth chip can be observed, thus our fit is likely not
the best fit for the object.  

\subsection{G347.0775$-$00.3927}

\citet{wheelwright_2010} find a similar sized disc to our best fitting
model, from 0.5--4\,au but with a lower inclination of $30\degr$.
However, our inclination of $84\degr$ is not well defined, and this does
lie within our lower error bound.  It should also be noted that the
bolometric luminosity from the RMS database used in their study has
since been revised to a lower value, which we use here.

\subsection{IRAS 08576$-$4334}

IRAS 08576$-$4334 has been extensively studied in recent years.
\citet{bik_2004} model the CO emission using an isothermal disc from
0.2--3.6\,au, with an inclination angle of $27\degr$, a CO number
density of $3.9\times 10^{21}\,\mathrm{cm}^{-2}$ and an excitation
temperature of $1600\,$K.  \citet{wheelwright_2010} used a similar
method, but utilise a disc model with fixed power laws, and show the
data to be well fit with a 0.09--0.78\,au disc, at an inclination of
$18\degr$, with a CO number density of $7.9\times
10^{21}\,\mathrm{cm}^{-2}$ and an intrinsic linewidth of $20\,$\kms.

\smallskip

\citet{ellerbroek_2011} find double peaked He\,{\sc i} and Fe\,{\sc i}
emission lines, with a separation of 60--100\,\kms, which they
conclude must originate from a circumstellar disc.  They suggest IRAS
08576$-$4334 is likely an intermediate mass YSO with a mass accretion
rate of $10^{-6}$--$10^{-5}\,$\msunyr, obtained from the determination
of the outflow mass loss rate.

\smallskip

In contrast to \citet{bik_2004} and \citet{wheelwright_2010}, we find
our observations are best fitted with larger disc from 0.6--6.5\,au, at
an inclination of $65\degr$.  We note that data of
\citet{wheelwright_2010} does not have sufficient wavelength range to
observe double-peaked emission lines beyond $2.297\,\micron$, and that
the lower resolution of the \citet{bik_2004} data may mask the
presence of these features, especially as they are not well defined
and asymmetric in our data.  Our best fitting model possesses a double
peak width of approximately $50\,$\kms, similar to that of the
He\,{\sc i} and metal emission lines in \citet{ellerbroek_2011}.

\smallskip

We note that the rotationally broadened lines from
2.306--2.309$\,\micron$ show asymmetry, with a depletion on the blue
side.  These transitions likely correspond to the cooler material
further out in the disc, and as such may be evidence for asymmetry in
the disc, which cannot be fitted with our axisymmetric disc model.

\subsection{IRAS 16164$-$5046}

\citet{bik_2004} model the CO emission of IRAS 16164$-$5046 using an
isothermal disc from 3.1--3.2\,au (with large errors) at an
inclination of $30\degr$, with an excitation temperature of $4480\,$K
and a number density of $4\times10^{20}\mathrm{cm}^{-2}$.  The extent
of the disc is consistent with our results within errors, however we
find a best fit that is closer to the central star, which would
account for our larger inner number density.  Our disc inclination of
$53\degr$ is higher.

\smallskip

\citet{bik_2006} find CO emission and Pfund series emission with a
width comparable to that of the CO emission, suggesting a common
kinematic origin.  However, due to the different conditions required
for both emission, they suggest the CO emission comes from the
midplane of a disc, while the Pf-lines comes from the ionised upper
layers.  They also note CO absorption from 2.33--2.35$\,\micron$,
indicative of cold, foreground molecular gas.

\subsection{M8E-IR}

\citet{wheelwright_2010} were able to fit the emission of this object
with a disc from 0.3--2.6\,au at an inclination of $16\degr$, with an
inner number density of $1 \times 10^{23}$\,\cm2 and a linewidth of
7\,\kms.  \citet{linz_2009} find an inclination of $18.5\degr$ and a
density exponent of $q = -2.05$ using $\alpha = 0.013$.

\smallskip

Our fits to M8E-IR did not satisfactorily converge using many starting
positions across the initial parameter space.  The spectrum was highly
reddened, and the fit suffered from several minima with similar
$\chi^{2}_{\mathrm{r}}$ values.  The only change applied between these
fits was the level of continuum subtraction applied.  There were two
issues with the data.  Firstly, the level of the chip 4 flux seemed
too high to be reproduced by the model, meaning that even solutions
that reproduced the rotational line structure were assigned poor
$\chi^{2}_{\mathrm{r}}$ values.  Secondly, the model was unable to
reproduce the relatively narrow structures on the red side of the
bandhead edge, which may be noise.  Because these features were across
a larger range of wavelength than the three broader lines at the edge
of chip 3, the fitting routine assigned them a higher weight which
resulted in poor fits to the bandhead slope and the broader lines.

\smallskip

To address this, we restricted the range of fitting to exclude this
region (as can be seen in Figure \ref{fig:results2}), and increased
the allowed upper linewidth to 60\,\kms, which produced a better fit
to the data, and produced similar best fitting parameters similar to
those in \citet{linz_2009}.

\section{Non-MYSO sources \& objects that were not fitted}
\label{sec:others}

\begin{figure}
\centering
\includegraphics[width=\columnwidth]{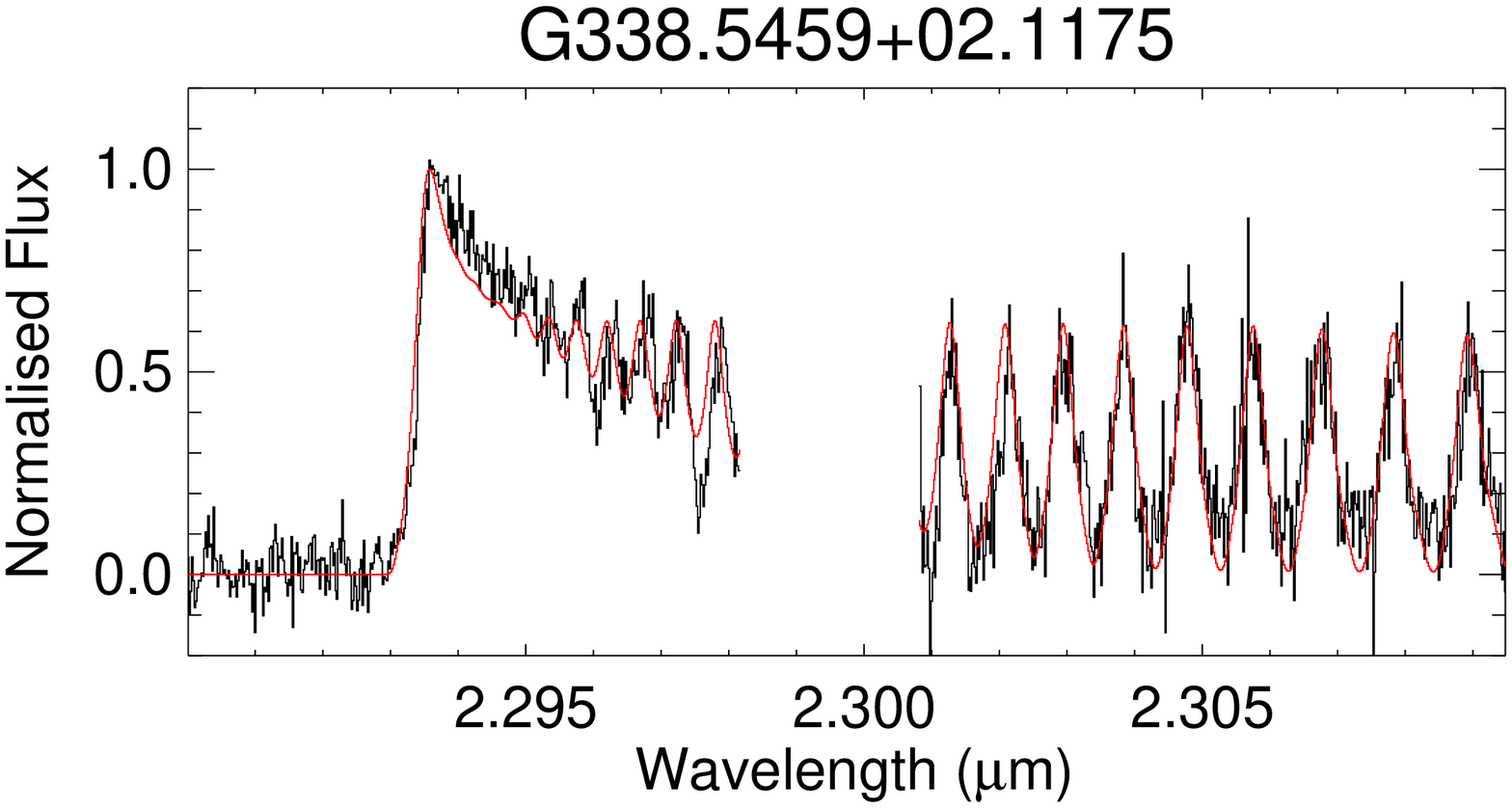}
\includegraphics[width=\columnwidth]{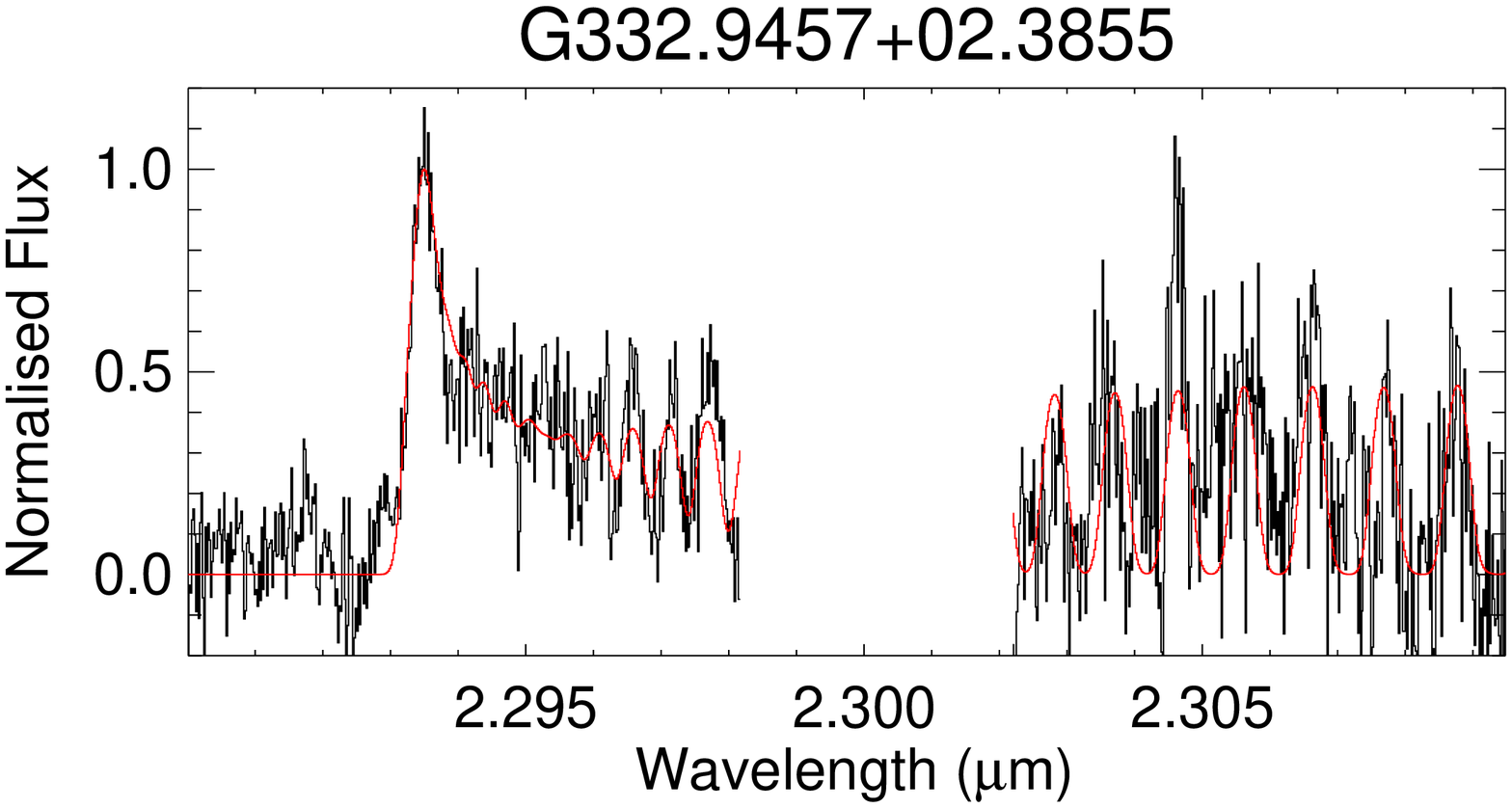}
\caption{Objects that were later confirmed to be non-MYSO sources, but
possess strong CO emission and were fitted with our disc model.}
\label{fig:nonmyso_results}
\end{figure}

The objects G332.9457$+$02.3855 and G338.5459$+$02.1175 were
originally thought to be MYSOs at the time of observing, but
subsequent determination of their bolometric luminosity has shown the
objects too faint for this to be the case, and they are likely lower
mass young stellar objects. We could not base our estimation of their
stellar parameters on the main sequence relationships, so we have
estimated their stellar parameters at the values shown in Table
\ref{tab:results}, which lie in the range of typical T Tauri star
values.  Changes to these parameters within these ranges had little
effect on the final fits.  The fits are presented in Figure
\ref{fig:nonmyso_results}.

\smallskip

Magnetically channelled accretion funnels have been suggested as a
possible source for CO bandhead emission in T Tauri stars
\citep{martin_1997}.  We obtain good fits to the data of two young
stellar objects using a simple disc model.  The sizes of the discs are
beyond the typical co-rotation radii for these objects.  In addition,
the intrinsic linewidths of both objects are similar to those of the
massive YSOs.  This suggests that the emission originates from
circumstellar discs regardless of the mass of the central YSO, and not
from accretion funnels.

\smallskip

\begin{figure}
\centering\includegraphics[width=\columnwidth]{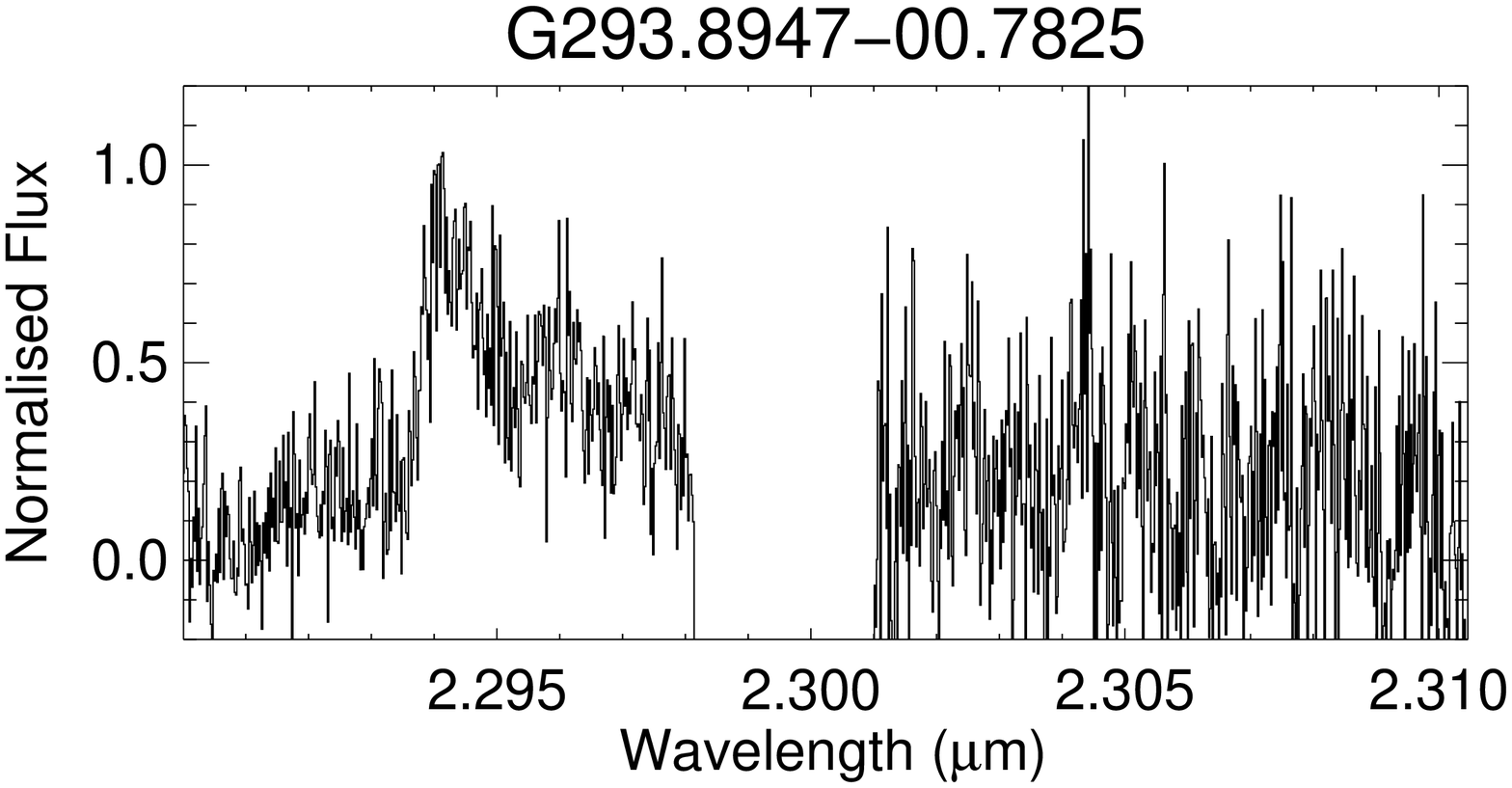}
\centering\includegraphics[width=\columnwidth]{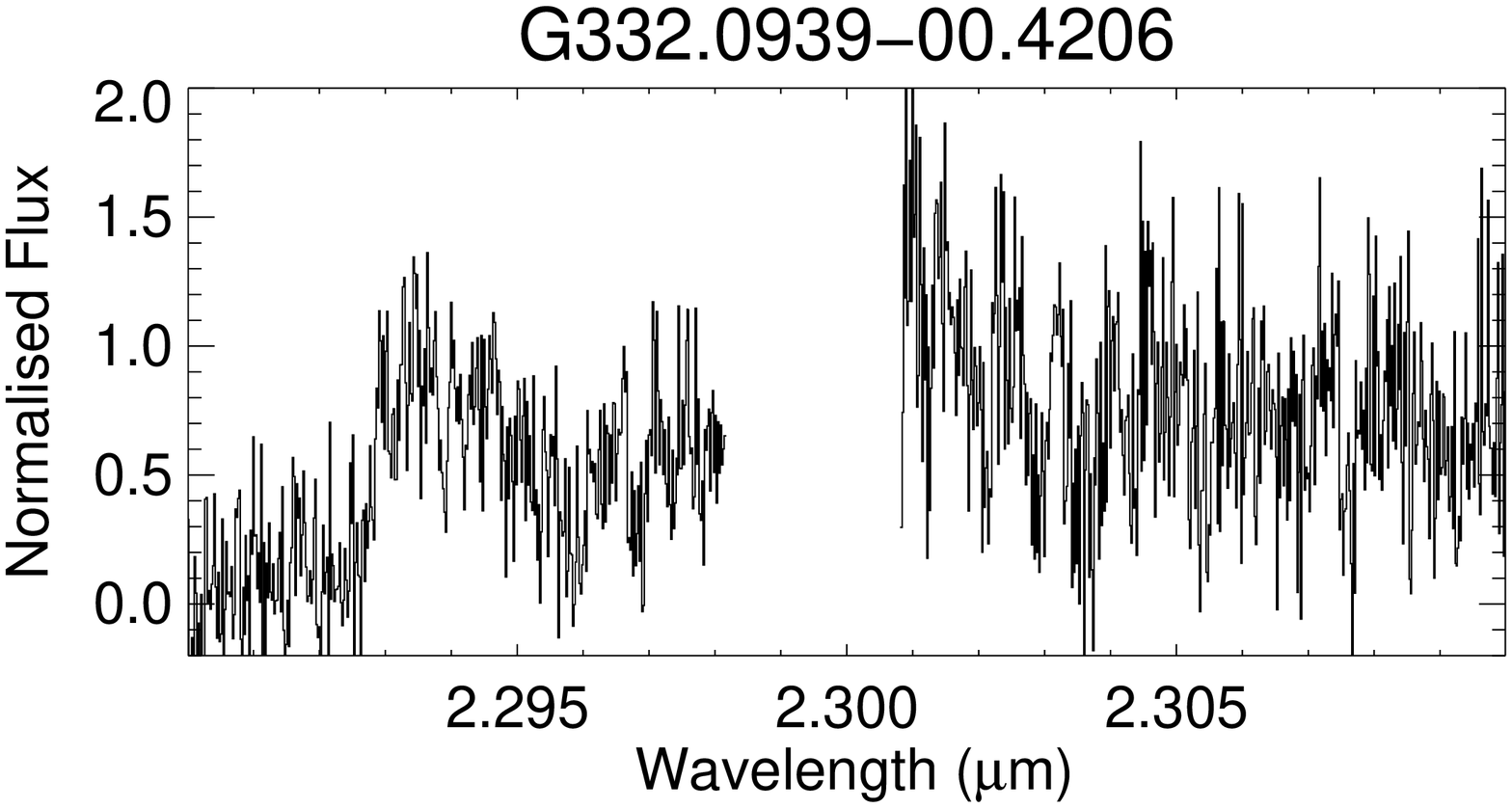}
\centering\includegraphics[width=\columnwidth]{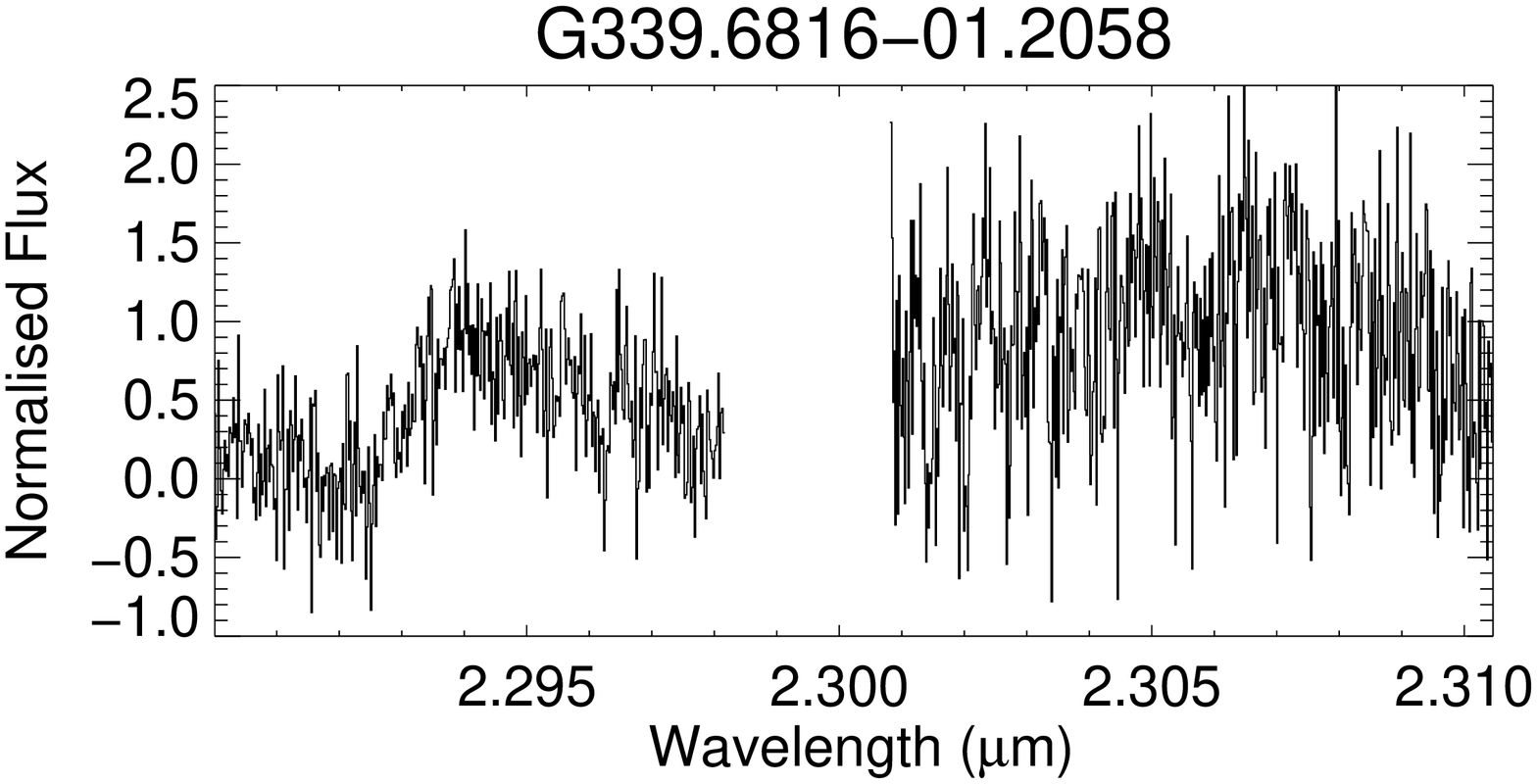}
\caption{Objects that possessed CO emission that was too weak for an accurate fit to be obtained.}
\label{fig:results_nofit}
\end{figure}

Finally, we note that three objects in the sample had CO emission that
was too weak, or a signal-to-noise ratio that was too low for an
accurate fit to be obtained.  For completeness, we include their
spectra in Figure \ref{fig:results_nofit}.

\bsp

\label{lastpage}

\end{document}